\documentclass{article}

\usepackage{PRIMEarxiv}

\usepackage[utf8]{inputenc} 
\usepackage[T1]{fontenc}    
\usepackage{hyperref}       
\usepackage{url}            
\usepackage{booktabs}       
\usepackage{amsfonts}       
\usepackage{nicefrac}       
\usepackage{microtype}      
\usepackage{lipsum}
\usepackage{fancyhdr}       
\usepackage{graphicx}%
\usepackage{multirow}%
\usepackage{amsmath,amssymb,amsfonts}%
\usepackage{amsthm}%
\usepackage{mathrsfs}%
\usepackage[title]{appendix}%
\usepackage{xcolor}%
\usepackage{textcomp}%
\usepackage{manyfoot}%
\usepackage{booktabs}%
\usepackage{algorithm}%
\usepackage{algorithmicx}%
\usepackage[noend]{algpseudocode}
\usepackage{listings}%

\usepackage{hyperref}     
\usepackage{nicefrac}     
\usepackage{microtype}   
\usepackage{lipsum}
\usepackage{fancyhdr}
\usepackage{wrapfig}
\usepackage{caption}
\usepackage{multirow}
\usepackage{bbm}
\usepackage{mathtools}
\usepackage{mathrsfs}
\usepackage{subcaption}
\usepackage{comment}
\usepackage{outlines}
\usepackage{rotating}
\newtheorem{definition}{Definition}
\usepackage{lineno}
\usepackage{natbib}

\usepackage{array}

\DeclareMathOperator*{\argmin}{arg\,min}

\title{The Value of AI Advice: Personalized and Value-Maximizing AI Advisors Are Necessary to Reliably Benefit Experts and Organizations

}

\author{
  Nicholas Wolczynski  \\
  University of Texas at Austin \\
  \texttt{nicholas@mccombs.utexas.edu} \\
   \And
  Maytal Saar-Tsechansky \\
  University of Texas at Austin \\
  \texttt{maytal@mail.utexas.edu} \\
   \And
  Tong Wang \\
  Yale University \\
  \texttt{tong.wang.tw687@yale.edu} \\
}

\begin{document}
\maketitle

\begin{abstract}
Despite advances in AI’s performance and interpretability, AI advisors can undermine experts' decisions and increase the time and effort experts must invest to make decisions. Consequently, AI systems deployed in high-stakes settings often fail to consistently add value across experts and organizations and can even diminish the value that experts alone provide. Beyond harm in specific domains, such outcomes impede progress in research and practice, underscoring the need to understand when and why different AI advisors add or diminish value. To bridge this gap, we stress the importance of assessing the value AI advice brings to real-world contexts when designing and evaluating AI advisors. Building on this perspective, we characterize key pillars -- pathways through which AI advice impacts value -- and develop a framework that incorporates these pillars to create reliable, personalized, and value-adding advisors. Our results highlight the need for value-driven development of AI advisors that advise selectively, are tailored to experts' unique behaviors, and are optimized for context-specific trade-offs between decision improvements and advising costs. They also reveal how the lack of inclusion of these pillars in the design of AI advising systems may be contributing to the failures observed in practical applications.
\end{abstract}



\section{Introduction}\label{sec:Intro}

Artificial intelligence (AI) systems are poised to play an increasingly vital role in augmenting human decision-makers across high-stakes environments, such as in healthcare \citep{pnas_occupation_2024}. Yet, despite advances in AI's generalization performance and interpretability, AI systems deployed to advise decision-makers in critical settings fail to consistently improve outcomes across contexts and expert behaviors \citep{jama_2024, Heterogeneity_Salz_Agarwal_Rajpurkar_2024,green_disparate_2019, lebovitz_engage_2022, green_algorithmic_2021, Jacobs2021-zs, Rezazade_Mehrizi2023-ia}. For example, AI models used to advise radiologists on diagnostic tasks \citep{lebovitz_engage_2022, Rezazade_Mehrizi2023-ia, Heterogeneity_Salz_Agarwal_Rajpurkar_2024} and having generalization accuracy comparable to or exceeding that of human experts have failed to consistently add value, sometimes leading experts to make poorer decisions or prolonging the decision-making process without adding sufficient value \citep{lebovitz_engage_2022, shinradio2023, Heterogeneity_Salz_Agarwal_Rajpurkar_2024}. These outcomes have also contributed to the under-utilization or abandonment of the tools after deployment \citep{lebovitz_engage_2022}. 
More broadly, a recent survey found that 56 papers on AI-assisted decision-making published between January 2020 and June 2023  reported that AI-assisted humans produced worse outcomes relative to what they were capable of alone \citep{Vaccaro2024}. Such detrimental outcomes not only run the risk of reducing trust in AI advice but impeding progress towards AI's positive impact and adoption in critical decision-making domains.

Despite AI advising systems' failures to consistently add value in practice, prior work on methods for producing AI advisors has not offered a comprehensive framework for producing AI advising systems that reliably add value and mitigate losses seen in practice. Such a framework is crucial to guide the design and evaluation of reliable, value-adding AI advisors as well as to facilitate AI's positive impact in an array of critical decision-making domains. This paper seeks to address this gap.

\begin{figure}
    \centering
    \includegraphics[width=0.55\textwidth]{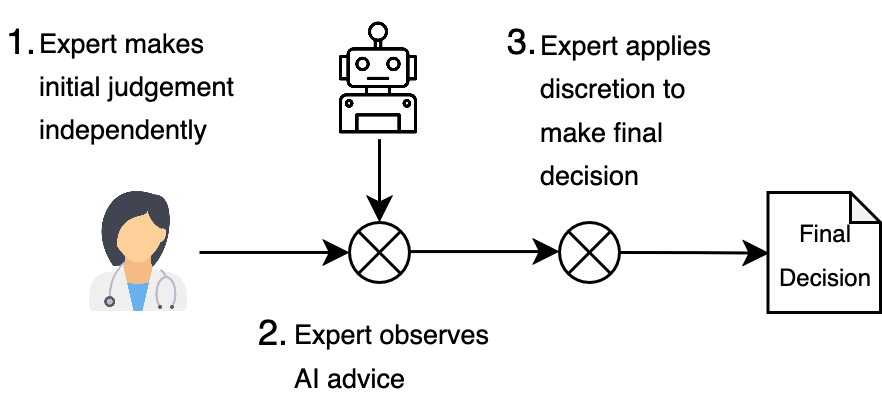}
    \caption{AI-Advised Decision Making}
    \label{fig:dec_sequence_intro}
\end{figure}

We focus on high-stakes settings that follow an AI-advised decision-making paradigm common in practice, shown in Figure \ref{fig:dec_sequence_intro}, where an expert first makes an independent judgment on the best decision to take, and an AI subsequently \emph{advises} the expert on the decision before the expert finalizes their decision \citep{lebovitz_engage_2022, GOMEZ2023102977,Goh2025, Yin2025, when_advise_2023, yin2019, riverain2022clearread}.\footnote{As an example, \href{https://radiology.healthairegister.com/products/riverain-clearread-xray-detect/}{ClearRead Xray}, used by hundreds of institutions in 13 countries, offers diagnostic advice to radiologists after the physician has performed an initial interpretation of the radiograph.} For brevity, throughout this paper we refer to such settings as AI-advised Decision-Making (AIaDM) settings and we refer to the AI in AIaDM settings as the \emph{AI advisor}. We focus on such settings because of their practical impact; however, the process and principles that we propose are general and can be adapted to produce value-based and reliable AI partners across contexts with different forms of human-AI collaboration.


We propose that enhancing the positive impact of AI advisors in practice requires a shift towards a \emph{value-based} approach to guide AI advisors' design, learning, and evaluation. Foundational to the framework we develop is a design that builds on opportunities to produce AI advisors that add the most value to their contexts and that simultaneously mitigate the losses they might introduce. Despite the failures in practice, prior research on the development \citep{pnas_ecg, Nishida2023, Abbas2024} and evaluation \citep{Heterogeneity_Salz_Agarwal_Rajpurkar_2024,NBERw31422,Vaccaro2024} of AI advisors has not proposed a framework for developing AI advisors that engage with and account for the different pathways through which  AI advice may heterogeneously either add or diminish value across contexts. Furthermore, while AI advisors in practice are trained to produce the most accurate decisions if the AI  were to make these decisions \emph{autonomously},  we consider settings in which the AI does not undertake decisions autonomously and may only advise a human, thus our design engages with how AI advice may impact the \emph{human's} decision outcomes. 




To our knowledge, this work is the first to (1) propose a value-based framework for producing AI advisors in AIaDM contexts, (2) empirically study its performance across diverse contexts relative to alternatives, (3) identify and enhance our understanding of contexts in which different AI advisors can add the most value or risk the greatest losses, and (4) uncover insights on the unique expertise and advising behaviors that AI advisors can give rise to and how they contribute to AI advisor's added value in their contexts. More concretely, our contributions are as follows:

\begin{enumerate}
    \item We first propose that the design objective of AI advisors is to \emph{reliably} offer the most \emph{added-value} to their context, which entails accounting for pathways through which AI advice can  impact  decision outcomes but also considering the costs  arising from the human's engagement with AI advice.
    \item Foundational to our contributions is the identification of key pathways — referred to as pillars — through which AI advice can impact the value added or diminished by an AI advisor and are thereby integral to achieving the AI advisor's goal. To the best of our knowledge, no prior work that developed AI advisors has proposed and motivated the benefits of simultaneously accounting for these pillars, and proposed methods for producing AI advisors that account for them so as to add the most value. Importantly, the pillars’ effects on decision-making outcomes are interdependent and thus introduce novel understandings of how AI advisors can add value or might lead to losses. 
    \item Building on the objective and pillars, we propose a framework for producing Reliable and Value-Maximizing AI (\textsc{ReV-AI}) advisors. \textsc{ReV-AI} advisors' expertise and advising behaviors are shaped by their context: they aim to complement the individual advisee's expertise and advice-taking behaviors, while accounting for the context's tradeoff between the costs and potential benefits from AI advice. \textsc{ReV-AI} advisors selectively offer advice and are specialized to learn and offer advice that is simultaneously sufficiently accurate and inherently convincing to the advisee so as to add the most value.  
    \item We use our framework to develop,  evaluate, and analyze a concrete proof-of-concept \textsc{ReV-AI} method -- referred to as TeamRules (\textsc{tr}) -- for generating inherently interpretable (\textsc{ReV-AI}) advisors. 
    \item Using real world data along with simulation grounded in human behavior, we explore the value that different AI advisors' can add or diminish across diverse contexts corresponding to different data domains, differential human decision-making abilities, advice-taking behaviors, and organizational trade-offs between decision benefits and costs. Our results reveal that across contexts, human expertise and advice-taking behaviors, \textsc{ReV-AI} advisors add significantly greater value and otherwise avoid diminishing the value offered by the human alone. Overall, our results demonstrate that producing AI advisors that can be relied on in practice to add the most value and avoid losses in high-stakes contexts requires that they offer effective responses by accounting for the pathways through which AI advice can add or diminish value. Consistent with evidence from practice, we also find that the most prevalent AI advisors in practice, not only may not add the most value, but introduce the greatest losses relative to the value already offered by the human alone. We also show why these outcomes may arise and reveal the contexts in which the most prevalent AI advisors can yield the greatest losses.
    
    \item We analyze two case studies grounded in real-world expert decision-making and advice-taking behavior to explore the expertise and advising strategies that \textsc{ReV-AI} advisors give rise to relative to alternatives. Our analysis reveals that \textsc{ReV-AI} advisors produce distinct expertise and advising behaviors that are shaped by their context and advisee and that these behaviors contribute directly to greater value creation and loss mitigation. 

\end{enumerate}

Our work advocates and aims to build the groundwork for value-based AI advisors in high-stakes settings. We hope that our work will inspire future research to advance AI's positive impact on high-stakes decisions and to develop other value-based AI frameworks for different forms of human-AI collaborations and contexts. Our work also aims to equip managers and stakeholders with a clearer understanding of the opportunity costs and limitations of current AI advisory systems and the contexts in which these limitations are likely to be most pronounced, while offering guidance to advance next-generation AI advisors that reliably enhance decision-making and consistently drive positive real-world outcomes.

\section{The Case for Reliable \& Value-Driven AI Advisors}\label{sec:theory}
In this section, we discuss the goal of AI advisors and build on prior work to introduce pathways through which advice impacts value. We then introduce a framework for developing AI advisors that reliably add value by accounting for these pathways. 

\subsection{The AI Advising Objectives: Offer Reliable, Value-Maximizing Advice} Perhaps the most fundamental objectives of an AI advisor are that (a) it learns to produce advice that \emph{maximizes the value it adds} to its context relative to what can be achieved by the \emph{human alone} and that (b) it is \emph{reliable}, such that when there is no opportunity to provide value-adding advice, it does not diminish the value delivered by the expert alone. Henceforth, we refer to advisors aiming to achieve these objectives as Reliable and Value-Maximizing AI (\textsc{ReV-AI}) advisors. Crucially for our work, the value added by an AI advisor encompasses not only the improvement in the decisions made by the human-AI team but also the practical costs associated with advising the human. Thus, developing AI advisors that effectively add value does not correspond to developing models that have high standalone performance on a given task; rather, AI advising models need to cost-effectively improve an \emph{expert's} decision outcomes and avoid diminishing them. Notably, this means that a worse-performing model that complements an individual's weaknesses while leveraging their strengths may be a better AI advisor to that expert than a model that has higher standalone performance across the decision space.   

\subsection{Pillars for Developing AI Advisors}
Central to our proposed value-based approach are pillars that characterize key pathways through which AI advice can add or diminish value in AIaDM settings and are thereby foundational to producing reliable and value-maximizing AI advisors.

\subsubsection{The overlooked costs of AI advice.}  
Recent studies on AI-advised classification tasks, such as medical diagnosis, often focus on optimizing the AI's standalone predictive performance \citep{pnas_ecg, Nishida2023, Abbas2024}. However, for AI advisors to add value in practice, improving predictive performance alone is not always sufficient. For instance, while double-reading diagnostic images by radiologists improves accuracy, it has not been widely adopted in the U.S. due to cost-effectiveness concerns  \citep{stefano_double_1995, Posso2016, pmid29594850}. Concretely, given double-readings double the diagnostic costs, its integration in practice hinges on whether the benefits in outcomes outweigh the costs. Consequently, in many key high-stakes contexts, for AI advisors to add value it is essential to account for any added costs incurred when AI advice is integrated into the decision-making process.

A significant cost, which has been largely overlooked in the development and evaluation of AI advisors, is the burden placed on experts when they engage with AI advice that conflicts with their independent judgment. AI advice can only improve an expert's independent judgment when it \emph{contradicts} it. Yet, evidence shows that decision-makers in high-stakes settings experience significant time costs when reconciling AI advice that contradicts their judgment \citep{Reverberi2022, Boyac2023, shinradio2023, lebovitz_engage_2022}. Similarly to previous innovations deemed cost-ineffective, excessive human engagement costs from reconciling disagreements with AI advice have been shown to impede the use of AI advice altogether  \citep{lebovitz_engage_2022,Lenharo2024}. 

Advancing AI advisors' positive impact on practice thus hinges on developing frameworks that produce \emph{cost-effective} AI advisors by accounting for the human costs of reconciling contradictory advice. While crucial to the value proposition of AI advisors in critical settings, the human engagement and its implications have not been considered thus far in practice or in research on AI advising methods. Importantly, the value added by AI advice depends on these costs, and the cost-effectiveness of advice is moderated by the trade-off within each context between added advising costs and the value produced from changes in the final decisions—which we henceforth refer to as the \emph{context's trade-off}. The framework we propose below accounts for the advising costs incurred to yield AI advisors that consistently add value. However, as we also demonstrate in the empirical evaluations, the framework generalizes to contexts which allow any added costs as long as the advice improves decision outcomes.

\subsubsection{Humans are idiosyncratic and imperfect advice-takers.}
\label{sec:imperf_ADB}
Human discretion in accepting or rejecting contradictory AI advice is both idiosyncratic and imperfect: individuals may reject accurate advice or accept incorrect recommendations \citep{dietvorst_algorithm_2015,green_disparate,green_principles,stop,bansal_does_2021,complete-me, Jacobs2021-zs, Rezazade_Mehrizi2023-ia, NBERw31422}. Prior work has shown that decision-makers tendency to accept and reject advice is informed by the human's own (and, possibly, miscalibrated) confidence as well as the AI's confidence in a given decision, both of which may vary across decision instances and differ across different decision-makers  \citep{will_you_accept,CHONG2022107018}. We refer to this tendency as an individual’s Algorithm Discretion Behavior (ADB). Prior work has also shown that humans' ADB is imperfect and they consequently do not necessarily benefit from AI advice even if the AI offers reliability measures such as confidence or is explainable \citep{Vaccaro2024, bansal_does_2021}. Importantly, ADB affects the cost-effectiveness of AI advice: given the human's engagement to reconcile contradictory AI advice is costly, a higher likelihood of rejecting AI advice reduces its cost-effectiveness. 

\subsubsection{Whether advice is accepted depends on the advice itself: offer inherently convincing advice.} 

An individual's likelihood of accepting AI advice varies across decision instances and depends on properties inherent to the advice, such as its confidence \citep{will_you_accept,CHONG2022107018, zhang_effect_of_conf}. Therefore, complementing AI recommendations with inherent properties of the advice has become crucial in high-risk settings, as such properties truthfully characterize the AI's recommendation \citep{laugel2019, rudin2019stop}. We refer to inherent properties of AI advice that impact the likelihood with which an individual will accept it as ADB-bound information. Crucial for learning cost-effective advice, research shows that the human's likelihood of accepting AI advice is influenced by the AI's reported confidence\footnote{While humans' likelihood of accepting AI advice is \emph{influenced by} AI's reported confidence, this influence does not necessarily lead to \emph{appropriate} likelihood of accepting AI advice. Thus, as mentioned in Section \ref{sec:imperf_ADB}, humans' may accept bad advice and reject good advice even if the AI offers reliability measures such as confidence.} in addition to the human's self-reported (and potentially miscalibrated) confidence in their own decision \citep{will_you_accept,CHONG2022107018, zhang_effect_of_conf}. Hence, developing value-maximizing AI advisors hinges on learning and offering advice that is simultaneously \emph{sufficiently} inherently persuasive and accurate to add value, per the context's trade-off. 

\subsubsection{Humans are idiosyncratic and imperfect decision-makers.} 
Lastly, the expert's decision-making behavior is central to the development of cost-effective AI advisors. Due to the idiosyncratic strengths and weaknesses of experts' judgment, the value of AI advice also depends on whether the AI's recommendation is more likely to be correct than the human's independent decision \citep{bansal_is_2021}. Importantly, given AI is an artifact that can be intentionally designed to add the most value,  different experts’ idiosyncratic strengths and weaknesses introduce opportunities to produce AI advisors with differential strengths to complement them.

\subsection{Framework for Producing Reliable Value-Maximizing AI Advisors}

\begin{figure*}[t!]
    \centering
    \makebox[0pt]{
    \includegraphics[width=0.95\textwidth]{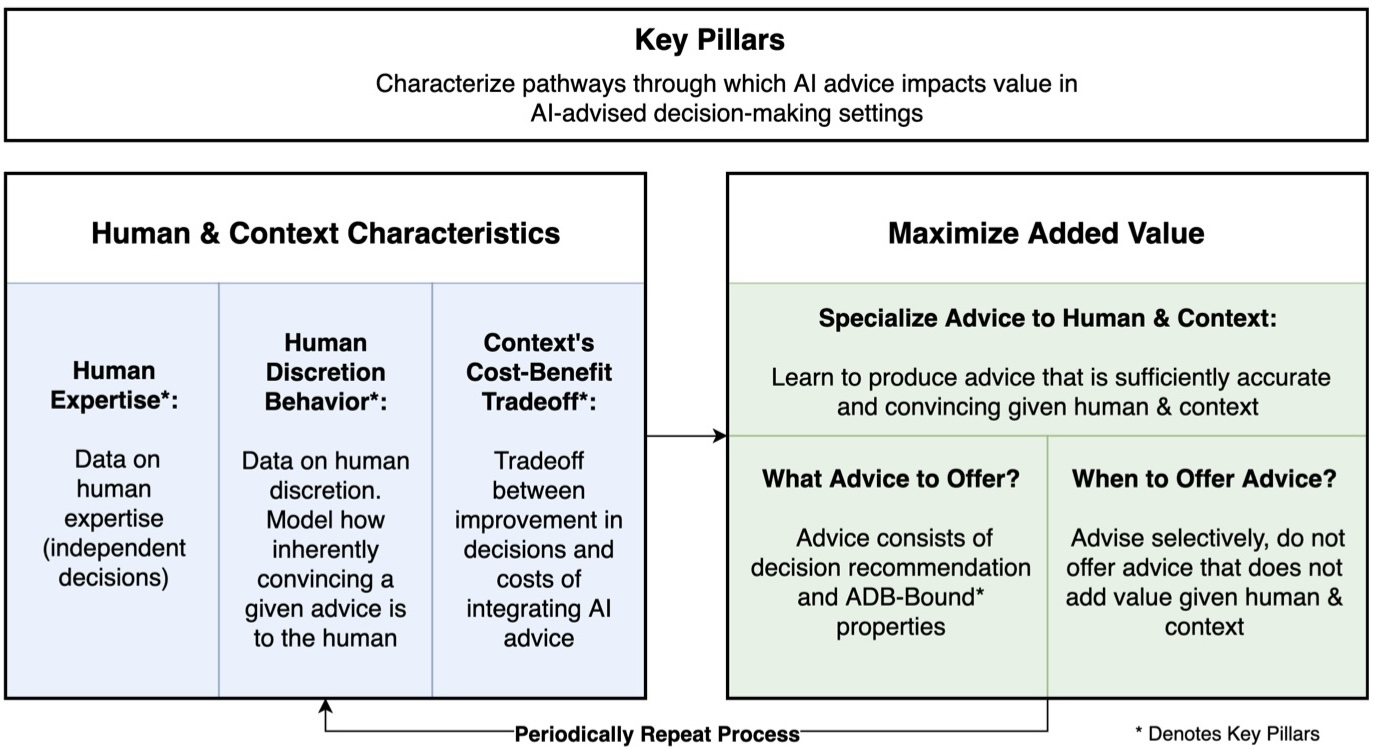}}
    \caption{Framework for producing reliable, value-maximizing AI advisors}
    \label{fig:frame}
\end{figure*}

We build on the pillars to develop a general framework for producing \textsc{ReV-AI} advisors. In a departure from existing AI advisors, the \textsc{ReV-AI} framework constitutes a value-based approach for developing and evaluating AI advisors. \textsc{ReV-AI} advisors selectively offer advice that maximizes \emph{added value} while avoiding actions that could diminish the value they provide to the \emph{expert} within their \emph{context}. Crucially, the advisors are grounded in the key pillars through which AI advice can impact the context’s value. Below we briefly note the key elements of the \textsc{ReV-AI} Learning Framework visualized in Figure \ref{fig:frame} and we formalize the framework in Section \ref{sec_probform}.

\textbf{Nature of advice: decision recommendation and inherent ADB-bound information.} A \textsc{ReV-AI} advisor provides both a decision recommendation and ADB-bound information that influences the likelihood of its advice being accepted by the human advisee. We consider advisors that provide confidence---the estimated likelihood that the advice is correct---as the complementary ADB-bound information offered. However, our framework can also be used to generate recommendations complemented by other inherent properties of AI advice that impact the human's ADB and, by extension, the value of the advice. We focus on confidence here because prior work has established how AI’s reported confidence along with the human’s self-reported confidence affect the human's receptiveness of advice \citep{will_you_accept}, allowing us to explore outcomes in diverse simulated AIaDM contexts. 

\textbf{Learning to offer value-maximizing advice personalized to a given human and context.} Our framework includes end-to-end learning of a model, which constitutes a repertoire of decision recommendations with corresponding confidence levels that are offered selectively during deployment, so as to add the most value across the decision space. Note that to add the most value to a context, a \textsc{ReV-AI} advisor can learn to produce advice that complements the human's expertise by adding significant value for some decisions that the human can benefit the most from at the possible expense of accuracy or persuasiveness on other decisions. 
\emph{Value-maximizing} advice is learned through an objective grounded in the pillars, which reflect how AI counsel either enhances or diminishes value in its context. The value of the advice is thus influenced by (1) its inherent persuasiveness given the human's confidence and corresponding ADB, (2) its likelihood of being correct relative to the likelihood the human's independent decision is correct, and (3) the context's trade-off between the human's engagement with the AI recommendation and the decision benefits of the advice. Hence, producing value-maximizing advice grounded in the \textsc{ReV-AI} advising objective requires data on the human's expertise, confidence, and ADB. We discuss data collection and related modeling in Section \ref{sec:methods}.


\textbf{Selective advising.} In some cases,  even the best advice learned by a \textsc{ReV-AI} advisor may not necessarily be cost-effective enough to add value. Thus, a crucial component of a \textsc{ReV-AI} advisor is a \emph{selective advising} mechanism that allows it to withhold advice when the best learned advice is expected to diminish value. This mechanism is integrated into the learning process, ensuring that the advice repository is optimized to add value while accounting for the selective withholding of advice if the advice is not expected to contribute positively.

\subsection{Problem Formulation}
\label{sec_probform}
We now formalize the framework for producing \textsc{ReV-AI} advisors shown in Figure \ref{fig:frame}. Key notation for this section is summarized in Table \ref{tab:notation}. Let $\mathcal{X}$ be the set of decision instances with labels $\mathcal{Y} = \{0,1\}$, where $\mathcal{X} \times \mathcal{Y} \sim {\mathcal{D}}$. A decision instance at index $i$ is represented by features $x_i$, and outcome $y_i$. We consider contexts reflecting common high-risk decision domains in which a human first makes an independent decision $h_i \in \{0,1\}$. The human is then advised by an AI advising model, denoted by $m$, that either withholds advice or offers advice in the form of a decision recommendation $\hat{y}_i$. The AI recommendation is complemented by the AI model's confidence\footnote{We consider the model's confidence because the impact of model confidence on human likelihood to accept advice has been studied in prior work. The framework can be instantiated with other forms of complementary ADB-bound information under the condition that the impact the ADB-Bound information has on ADB is modeled.} $c^M_i$ reflecting the AI's assessment of the likelihood that its advice is accurate. Finally, the human produces a final decision, by either accepting the recommendation $\hat{y}_i$ or selecting their own independent decision $h_i$. 

When the AI does not offer advice, the human's independent decision becomes the human-AI team's final decision; when the AI offers advice that contradicts the human's initial judgment, i.e., ${h}_i \neq \hat{y}_i$, the human applies their discretion before arriving at a final decision.  We denote with $a_i$ whether the human DM accepts the AI recommendation for instance $i$, where $a_i = 1$ denotes that the human DM accepts the AI advice, and $a_i = 0$ means the DM rejects it. The human's discretion over AI advice is represented by the likelihood of the human accepting the AI advice $p(a|c^M, c^H, \hat{y} \neq h)$, and is estimated with $\hat{p}(a|c^M, c^H, \hat{y} \neq h)$, where $c^H$ represents the human DM's confidence in their original decision  $h$. Throughout the rest of the paper, we refer to $\hat{p}(a|c^M, c^H, \hat{y} \neq h)$ as the human's estimated discretion model and denote it with $\hat{p}(a)$, for brevity.

\begin{table}[]
\centering
\small
\caption{Key Notation}
\label{tab:notation}
\begin{tabular}{p{0.2\textwidth} | p{0.65\textwidth}}
\toprule
\multicolumn{1}{l|}{\textbf{Notation}} &
  \textbf{Description} \\ \toprule
\multicolumn{1}{l|}{$x_i$} &
  Feature values for instance at index i. \\ \hline
\multicolumn{1}{l|}{$y_i$} &
  Ground truth label for instance at index i. \\ \hline
\multicolumn{1}{l|}{$h_i$} &
  Human expert's initial decision (before seeing AI advice) for instance at index i. \\ \hline
\multicolumn{1}{l|}{$c^H_i$} &
  Human expert's confidence in their initial decision for instance at index i. \\ \hline
\multicolumn{1}{l|}{$c^M_i$} &
  Model's reported confidence in its decision advice for instance at index i. \\ \hline
\multicolumn{1}{l|}{$\hat{y}_i$} &
  Model's decision advice for instance i. \\ \hline
\multicolumn{1}{l|}{$\hat{p}(a|c^M, c^H, h\neq\hat{y}))$, *$\hat{p}(a)$ for brevity} &
  Estimated model of human's algorithm discretion behavior (ADB) - likelihood human accepts a contradicting recommendation given their confidence in their initial decision and model's  confidence in its advice. \\ \hline
\multicolumn{1}{l|}{$\hat{p}(y|x)$} &
  Estimated model of ground truth label for an instance. \\ \hline
\multicolumn{1}{l|}{$R, R^+, R^-$} &
  Sets of rules. $R^+$ and $R^-$ denote sets of rules that correspond to a model producing positive and negative decision advice if covered by each rule set, respectively. \\ \hline
\multicolumn{1}{l|}{$\mathcal{L}(y, \hat{y}, h, \hat{p}(a), \alpha)$} &
  Expected team loss from one AI advice over human's likelihood to accept it \\ \hline
\multicolumn{1}{l|}{$\mathcal{V}(y,d)$} &
  Denotes the context-relevant loss incurred from the human's final decision $d$ given ground truth $y$. \\ \hline
\multicolumn{1}{l|}{$\alpha$} &
  Represents the context's advising cost-benefit trade-off. \\ \hline
\multicolumn{1}{l|}{$\mathbb{C}(R, x_i)$} &
  Rule coverage indicator. True if instance with feature values $x_i$ is covered by any rule in $R$. \\ \hline
\multicolumn{1}{l|}{$\psi(y_i, y^*_i, h_i, \hat{p}(a), \alpha)$} &
  Indicator function that determines whether a prospective advice $y^*$ adds value to the expert in expectation. \\ \hline
\multicolumn{1}{l|}{$\mathbb{I}\{\text{condition}\}$} &
  Indicator function. Returns $1$ if condition is met, $0$ otherwise. \\ \hline
\multicolumn{1}{l|}{TDL} &
  Team decision loss from final decisions (loss from decisions only, does not include advising costs). \\ \hline
\multicolumn{1}{l|}{TTL} &
  Total team loss from final decisions. \\ \bottomrule 
\end{tabular}
\end{table}

A \textsc{ReV-AI} advisor's advice $\hat{y}$ for a decision instance is either a decision recommendation $z \in \mathcal{Y}$ or the lack of a decision recommendation, such that the withholding of advice corresponds to recommending the human's initial judgment $h$. Formally, $\hat{y}$ is given by:  

\begin{align}
\label{eq:theo_dec_rule}
    \hat{y} &= \begin{cases}
        z \in \mathcal{Y}, & \text{if decision advice is offered} \\
        h, & \text{if advice is withheld}
    \end{cases}
\end{align} 

Simultaneously, the selective advising decision -- whether to provide advice and the specific choice of advice to provide $z$ -- is personalized to the human and context's cost-benefit trade-off, with the goal of  maximizing the advisor's added value:

\begin{equation} 
\label{eq:theo_obj}
  m^* = \underset{m \in \mathcal{M}}{\argmin} \bigg[\mathbb{E}_{x,y,h \sim \mathcal{D}} \bigg(\mathcal{L}(y, \hat{y}, h, \hat{p}(a), \alpha)\bigg)  \bigg]
\end{equation}

where $\mathbb{E}_{x,y,h \sim \mathcal{D}}$ denotes expectation over the input space $(X,Y)$ and the human's independent decisions over it with joint distribution $D$. $m^*$ is thus the model from a set of possible models $\mathcal{M}$ that minimizes $\mathbb{E}_{x,y,h \sim \mathcal{D}} (\mathcal{L}(y, \hat{y}, h, \hat{p}(a), \alpha))$. $\mathcal{L}$ is the \emph{Expected Team Loss} over the  human's discretion $a$ resulting from the AI advice $\hat{y}$, the decision instance $(x,y)$, and the human's independent decision $h$, given by: 

\begin{align}
\label{eq:obj}
\mathcal{L}(y, \hat{y}, h, \hat{p}&(a), \alpha) = \hat{p}(a)\mathcal{V}\big(y, \hat{y}\big) \\ \nonumber &\text{\indent} +  \big(1-\hat{p}(a)\big)\mathcal{V}(y, h) + \alpha\mathbb{I}\{\hat{y} \neq h\},
\end{align}

where $\mathcal{V}(y,d)$ denotes the context-relevant loss incurred from the human's final decision $d$ given ground truth $y$. $\mathcal{V}(y,d)$ can be selected to fit the context, and the framework makes no assumptions on the loss produced by different types of errors. $\alpha$ denotes the context's trade-off between reconciling contradictory advice and benefits from improving decisions, such that the loss from reconciling one contradictory advice is equivalent to the following loss from a decision error: $\frac{\alpha}{\mathcal{V}(y,d |y \neq d)}$. Equivalently, the trade-off implies that $\frac{\mathcal{V}(y,d |y \neq d)}{\alpha}$ contradicting pieces of advice produce loss equivalent to that of a single incorrect decision.   
Note that Equation (\ref{eq:obj}) reflects that when $m$ offers contradictory advice, i.e., when $\hat{y} \neq h$, the expected loss depends on the estimated likelihood $\hat{p}(a)$ that the human accepts the advice; the loss in this case also includes the cost $\alpha$ from the human's reconciliation of the contradiction. When $m$ does not provide decision advice or when it provides advice that agrees with the human's initial judgment $h$, then $\hat{y} = h$, and the loss becomes $\mathcal{V}(y,h)$. 

The total expected loss produced by an AI advisor, $\mathbb{E}_{x,y,h \sim \mathcal{D}} \bigg(\mathcal{L}(y, \hat{y}, h, \hat{p}(a), \alpha)\bigg)$, is henceforth referred to as the \emph{Total Team Loss} (TTL). Importantly, as reflected in our discussion and shown in Equation (\ref{eq:obj}), the loss includes the loss from reconciling contradictions in addition to the loss from incorrect final decisions. Ultimately, offering value-adding advice hinges on improvements achieved cost-effectively per the context's cost-benefit trade-off. 

The \textsc{ReV-AI} advisor is informed by the underlying data domain $\mathcal{X} \times \mathcal{Y}$, data reflecting the human's expertise, confidence, and ADB for estimating $p(a|c^M, c^H, \hat{y} \neq h)$, and the context's tolerance for added costs to improve decisions.  Data reflecting the human's expertise - consisting of the human's independent decisions $\mathcal{H} = \{0,1\}$ for instances in $\mathcal{X}$ - is often available in historical data and which otherwise can be acquired for the corresponding domain. Prior work also established how an individual's ADB -- i.e., likelihood of accepting a given piece of AI advice for a given decision, $\hat{p}(a)$ -- can be effectively estimated based on the human's self-reported (possibly miscalibrated) confidence in their decision and the AI's confidence, $c^H$ and $c^M$, respectively 
\citep{will_you_accept,CHONG2022107018}.

\subsection{Data Acquisition, Training, and Evaluation}
In most high-stakes contexts, regulation and best practices dictate that the efficacy and safety of new technologies should be evaluated prior to deployment to mitigate harms. We follow this paradigm and consider a flow in which the expert initially interacts with AI advice over time in a risk-free environment, prior to producing the AI advisor. During this phase, data is collected, after which a \textsc{ReV-AI} advisor is trained, and subsequently the efficacy and safety of the advisor are evaluated. Concretely, building on prior work \citep{will_you_accept, CHONG2022107018}, for a subset of risk-free training decision instances $(x,y)$, the human's decision $h$, self-reported confidence $c^H$, and decision to accept or reject the AI advice $a$, are acquired and used to model the human's ADB. AI advice is initially offered by a general-purpose model trained on the joint space $\mathcal{X} \times \mathcal{Y}$. The remaining set of risk-free training decision instances are used to train the \textsc{ReV-AI} advisor. Complete details of modeling the human's ADB is outlined in Appendix A.2.

\section{Methods: TeamRules, An Inherently Interpretable ReV-AI Advisor}\label{sec:methods}
\begin{figure*}[t!]
    \centering
    \makebox[0pt]{
    \includegraphics[width=0.95\textwidth]{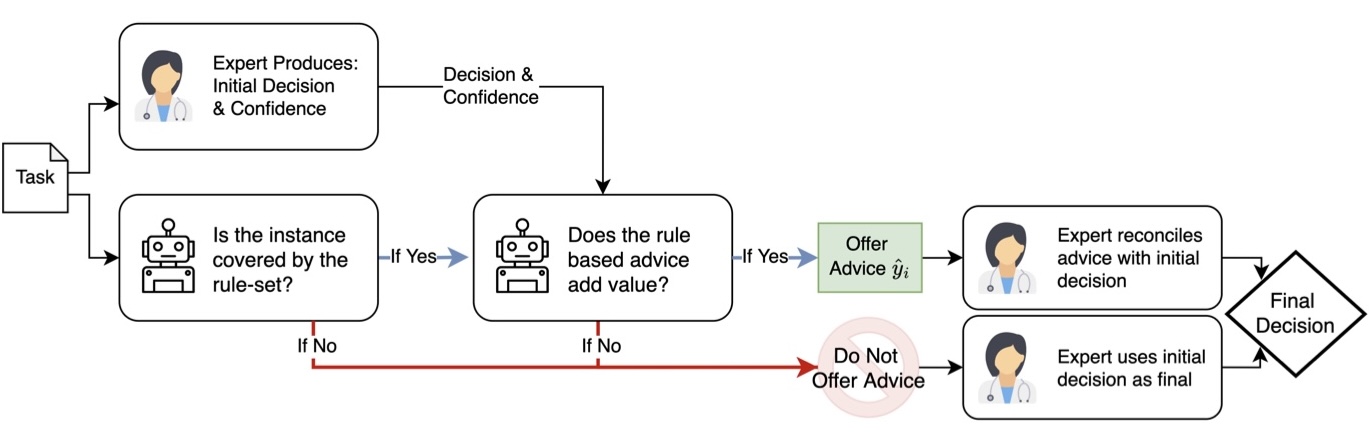}}
    \caption{TeamRules Advising Process in Deployment}
    \label{fig:deploy}
\end{figure*}

The \textsc{ReV-AI} learning framework can be used to derive different concrete methods for generating \textsc{ReV-AI} advisors of different model classes.  In this work, we develop and evaluate a rule-based prototype algorithm, TeamRules (\textsc{tr}), which produces \textsc{ReV-AI} advisors that generate decision recommendations in the form of inherently interpretable rules. Inherently interpretable models are especially suitable for and are commonly used to advise humans across high-stakes contexts \citep{rudin2019stop,wang2021scalable,rich,BALAGOPAL2021102101,edit_machine, pmlr-v139-biggs21a, Subramanian_Sun_Drissi_Ettl_2022, prescriptive_relu} because they faithfully represent the model’s actual decision-making process \citep{dangers_laugel}. Rule-based models also present challenges for optimization, hence we also show how these challenges can be effectively addressed. The \textsc{tr} advising process during deployment is shown in Figure \ref{fig:deploy}. 

Independently, as done in practice with non-stationary concepts, such as human preferences or choices and market outcomes, changes in the underlying concept merely entail degradation in the model's performance which can be monitored to trigger retraining and evaluation, as desirable. Importantly, our framework does not assume perfect estimates of ADB and we empirically demonstrate that producing effective \textsc{ReV-AI} advisors does not rely on perfect estimates. In the evaluations reported here we induce the discretion model from data and later study how increasing mis-estimates of the human's ADB impact the AI advisor's performance and benefits.



\subsection{Producing TeamRules}
Our framework can be used to derive different methods that generate \textsc{ReV-AI} advisors for different humans and contexts. We develop a rule-based prototype advisor, TeamRules (\textsc{tr}). Rules have been shown to offer an especially intuitive form of inherently interpretable advice for high-stakes decision-makers, given their transparent inner structures and expressiveness \citep{rudin2019stop,wang2021scalable,rich,BALAGOPAL2021102101,edit_machine}. The purpose of the prototype is to first demonstrate the use of the framework to derive a method for generating \textsc{ReV-AI} advisors. Most importantly, the prototype allows us to study and characterize how the behaviors of different \textsc{ReV-AI} advisors produced by (\textsc{tr}) vary across contexts, and to study the impact of \textsc{ReV-AI} advisors on value added to the human relative to that of alternatives. 

We begin by formalizing the terms \emph{rule} and \emph{rule set} used to advise a decision-maker:
\begin{definition}[Rule] 
A \emph{rule} $r$ is a logical expression made of conditional statements about a conjunction of a subset of feature values. $r$ \emph{covers} an example $x_i$, denoted as $\mathbb{C}(r, x_i) = 1$, if the logical expression evaluates to true given $x_i$. 
\end{definition}

\begin{definition}[Rule Set]
A \emph{rule set} $R$ is a collection of rules. $R$ \emph{covers} an example $x_i$, denoted $\mathbb{C}(R, x_i) = \mathbb{I}\{\sum_{r \in R} \mathbb{C}(r,x_i) \geq 1\}$, if at least one rule $r$ in $R$ covers $x_i$.
\end{definition} 

Given a training set of instances $D = \{x_i, y_i, h_i, c^H_i\}_1^n$, \textsc{tr} learns a set of rules $R = \{R^+, R^-\}$ to advise the human, such that rules in $R^+$ or $R^-$ can be used to advise 1 or 0 on instances that satisfy them, respectively. Note that a distinct property of rule-based advice is that a given rule set $R$ can be produced so that it does not cover all instances. Following our framework, to minimize the expected loss from the human's final decisions, a \textsc{tr} advisor \emph{selectively} offers advice for an instance $x_i$ if $x_i$ is covered by a rule in $R$ (i.e., if $\mathbb{C}(R, x_i)=1$) and if the expected Total Team Loss (TTL) from offering the advice is lower than the expected TTL when advice is withheld. In summary, Figure \ref{fig:deploy} demonstrates \textsc{TR}'s  advising process which consists first of providing possible advice determined by the rule set $R$ and by further assessing the expected added value of advice when an instance is covered by $R$.   

The learned rule set $R$ simultaneously determines (a) for which \emph{ decision instances} advice \emph{may} be offered, and (b) \emph{what} rule-based advice can be offered to the human so that it minimizes the TTL.  Below we outline the optimization method for deriving $R$ that minimizes the TTL from the human's final decisions by accounting for the human's expertise, ADB, and the context's cost-benefit trade-off. 

\textbf{Selective Advising.} Formally, \textsc{tr}'s selective advising at inference time is outlined in Equations (\ref{eq:dec_rule}) and (\ref{eq:withhold}). For any decision instance represented by feature values $x_i$ encountered during deployment, \textsc{tr} either offers advice or withholds advice which is equivalent to recommending the human's independent decision $h_i$, given by:\footnote{Note how Equation \ref{eq:theo_dec_rule} from the \textsc{ReV-AI} advising framework is concretely operationalized to form \textsc{tr}'s rule-based advising process shown in Equation \ref{eq:dec_rule}.}
\small
\begin{equation} 
\label{eq:dec_rule}
 \hat{y}_i = 
   \begin{cases}
    1, &  \text{if } \mathbb{C}(R^+, x_i) \wedge \psi(y, 1, h_i, \hat{p}(a), \alpha)  \\
    0, &  \text{if }  \mathbb{C}(R^-, x_i) \wedge \neg \mathbb{C}(R^+, x_i) \wedge \psi(y, 0, h_i, \hat{p}(a), \alpha)\\
    h_i, & \text{otherwise}
\end{cases} 
\end{equation}
\normalsize

Where $\psi(y, y^*, h, \hat{p}(a), \alpha)$ shown below is the indicator function that determines whether a prospective advice $y^*$ adds value in expectation, i.e., whether the expected loss from the advice ($\mathbb{E}_{y}[\mathcal{L}(y,  y^*_i, h_i, \hat{p}(a), \alpha)]$) is lower than the expected loss from withholding advice ($\mathbb{E}_{y}[\mathcal{L}(y, h_i, h_i, \hat{p}(a), \alpha)]$: 
\begin{align}
\label{eq:withhold}
    \nonumber \psi(y_i, y^*_i, &h_i, \hat{p}(a), \alpha) = \\& \mathbb{I}\{\mathbb{E}_{y}[\mathcal{L}(y,  y^*_i, h_i, \hat{p}(a), \alpha)] \\& \nonumber < \mathbb{E}_{y}[\mathcal{L}(y, h_i, h_i, \hat{p}(a), \alpha)]\}
\end{align}

To take expectation over the ground truth labels, $\mathbb{E}_{y}$, we train a model $\hat{p}(y|x)$ using an XGBoost classifier \citep{xgboost} with hyperparameters selected using grid-search cross-validation \citep{scikit-learn}. In addition, the total team loss from the human's final decision for decision instance at index $i$, $\mathcal{L}(y_i, \hat{y_i}, h_i, \hat{p}(a), \alpha)$, is shown in Equation (\ref{eq:obj}). In our implementation of \textsc{tr}, we set the instance-level decision loss in Equation (\ref{eq:obj}) to: $\mathcal{V}(y, d) = \lambda_0^{(1-y)}\lambda_1^{(y)}\mathbb{I}(y \neq d)$, 
where $\lambda_0$ and $\lambda_1$ are the decision losses incurred from false positives and false negatives, respectively.  

\textsc{tr} accompanies advice with ADB-bound information, namely, its corresponding confidence $c^M_i$, which is the precision of the rule used to produce the advice. When an instance is covered by multiple rules producing the same outcome, the highest precision rule is used. Specifically, the confidence is given by: 

\begin{equation}
\label{conf_rule}
\begin{aligned} 
&c^M_i = \\&  \underset{r \in R^{sgn(\hat{y}_i)} : \mathbb{C}(r,x_i) = 1}{\max}\Bigg[\frac{\sum_{j=1}^{n}\big(\mathbb{I}\{\mathbb{C}(r, x_j) \wedge \mathbb{I}(y_j = \hat{y}_i)\}\big)}{\sum_{k=1}^{n}\mathbb{C}(r, x_k)}\Bigg] 
\end{aligned}
\end{equation}

Note that the confidence provided is an inherent characterization of the quality of the rule used to provide the piece of advice. 

\begin{algorithm*}[t]
\scriptsize
\caption{\textsc{TR} Rules Generator}\label{alg:team}
\begin{algorithmic}[1]
\State \textbf{Input:} Data $(x_i, h_i, c^H_i, y_i) \text{ }\forall i \in \{0,...,n\}$, Discretion Model $\hat{p}(a)$
\State \textbf{Parameters: $T, \alpha, C_0, \beta_0, \beta_1, \beta_2$} 
\State $\Gamma^+ \gets \text{RF-Generated Positive Candidate Rule Set}$ 
\State $\Gamma^- \gets \text{RF-Generated Negative Candidate Rule Set}$
\State $R_0 \gets \emptyset$, $R^* \gets R_0$
\For{$t=1...T$}
\State $R_t \gets R_{t-1}$
\For{$i=1...n$}
\State $\text{Produce }\hat{y}_i$, $c^M_i$  \Comment{According to Equations (\ref{eq:dec_rule}) and (\ref{conf_rule}).}
\EndFor
\State $\epsilon \gets $ draw instance at index $i$ with probability: $\frac{\mathcal{L}(y_i, \hat{y}_i, h_i, \hat{p}(a_i | c^M_i, c^H_i, \hat{y}_i \neq h_i),\alpha) - \alpha\mathbb{I}\{\mathcal{V}(y_i, 1-y_i)\hat{p}(a_i|c^M_i, c^H_i, \hat{y}_i \neq h_i) > \alpha\}}{\sum_{j=1}^n\bigg(\mathcal{L}(y_j, \hat{y}_j, h_j, \hat{p}(a_j | c^M_j, c^H_j, \hat{y}_j \neq h_j),\alpha) - \alpha\mathbb{I}\{\mathcal{V}(y_j, 1-y_j)\hat{p}(a_j|c^M_j, c^H_j, \hat{y}_j \neq h_j) > \alpha\}\bigg) }$  
\If{($\mathbb{C}(R_{t-1},x_\epsilon)$ and ($\hat{y}_\epsilon \neq y_\epsilon$)) or (($h_\epsilon \neq \hat{y}_\epsilon$) and ($\hat{p}(a_i|c^M_i, c^H_i, \hat{y}_i \neq h_i) < \alpha$))}
\If{$(y_{\epsilon} = 0)$}
\State $R_{t} \gets$ cut or replace rule from $R^+_t$
that covers $x_\epsilon$
\Else{}
\State $R_{t} \gets$ cut or replace rule from $R^-_t$ that covers $x_\epsilon$
\EndIf
\Else{}
\State $R_{t} \gets$ add rule to $R^{sign(y_\epsilon)}_{t}$
\EndIf
\For{$i=1...n$}
\State $\text{Produce }\hat{y}^*_i$, $c^{*M}_i$ \Comment{Using updated rule-set $R_t$}
\EndFor
\If{$\sum_{i=1}^n\mathcal{L}(y_i, \hat{y}_i^*, h_i, \hat{p}(a_i | c^{*M}_i, c^H_i, \hat{y}_i^* \neq h_i), \alpha) < \sum_{i=1}^n\mathcal{L}(y_i, \hat{y}_i, h_i, \hat{p}(a_i | c^M_i, c^H_i, \hat{y}_i \neq h_i), \alpha)$}
\State $R^* \gets R_{t}$
\EndIf
\If{$\exp \left( {\frac{\sum_{i=1}^n\mathcal{L}(y_i, \hat{y}_i, h_i, \hat{p}(a_i | c^M_i, c^H_i, \hat{y}_i \neq h_i), \alpha) - \sum_{i=1}^n\mathcal{L}(y_i, \hat{y}^*_i, h_i, \hat{p}(a_i | c^{*M}_i, c^H_i, \hat{y}^*_i \neq h_i), \alpha)}{C_0^{\frac{t}{T}}}} \right) \leq$  random$([0,1])$} 
\State $R_t \gets R_{t-1}$ 
\EndIf
\EndFor
\State \textbf{Output:} $R^*$
\end{algorithmic}
\end{algorithm*}

\subsection{Optimization Approach}

We develop a training algorithm based on the core idea of simulated annealing to learn a rule set $R$ that minimizes the team's loss in Equation (\ref{eq:obj}). Simulated annealing is well-suited for  our discrete space, and has been used successfully in previous rule-based models  \citep{sim_anneal, wang2017bayesian}. The complete algorithm is shown in Algorithm \ref{alg:team}.  The optimization procedure takes as input the set of training instances $D = \{x_i, y_i, h_i, c^H_i\}_1^n$, along with the cost of reconciling a piece of contradictory advice $\alpha$ and the human discretion model $\hat{p}(a)$.  Initially, a set $\Gamma$ of $\beta_0$ candidate rules is generated using an off-the-shelf rule mining algorithm, such as FP-growth \citep{wang_gaining_nodate}. We follow prior work and established practices to set the hyperparameters which we discuss in Appendix A.3.  
The initial rule set $R_o$ is set to an empty set so that decisions are made exclusively by the human. At each subsequent iteration $t$,  a new rule set is proposed by either adding or dropping a rule to the current solution. This new set is accepted with a probability that depends on the quality of the proposed solution — i.e., the lower the total team loss (TTL) across all training instances, the higher the likelihood of acceptance. The probability also depends on the temperature, which decreases over iterations — i.e., the higher the temperature, the greater the likelihood of acceptance. As a result, the algorithm explores more aggressively at the beginning of the search, being more likely to accept solutions even if they are worse than the current one to thoroughly explore the space. Towards the end, it becomes more conservative, accepting only solutions that improve the existing rule set. This approach allows the algorithm to progressively minimize the TTL while avoiding getting stuck in local minima. 

By following the principle of stochastic local search \citep{hoos2018stochastic}, it is possible to explore changes to R by either adding to or removing from $R$ a randomly selected rule. However, to expedite convergence, revisions to $R$ are guided by identifying instances that contribute to the loss and exploring changes that may decrease this loss. Specifically, at each iteration, an instance with index $\epsilon$ is drawn at random with the probability of being drawn proportional to its relative contribution to the empirical TTL over all instances, as shown in Line 9 Algorithm \ref{alg:team}. Note, however, that the advising loss is ignored from the loss contribution if the possible improvement in  decision loss $\mathcal{V}(y_i, 1-y_i)\hat{p}(a_i|c^M_i, c^H_i, \hat{y}_i \neq h_i)$ is greater than the incurred advising cost $\alpha$. We do not account for the advising cost in such cases because the benefit from attempting to change the human's decision is greater than the cost incurred from reconciling, so we do not treat such examples as needing a correction, despite them contributing to the final TTL. Once an instance is selected, $R_t$ may be adapted by either adding, removing, or replacing a rule if the adaptation decreases the empirical TTL. Subsequently, the change is accepted only if it reduces the empirical TTL, i.e., if $\sum_{i=1}^n\mathcal{L}(y_i, \hat{y^*}_i, h_i, \hat{p}(a), \alpha) < \sum_{i=1}^n\mathcal{L}(y_i, \hat{y_i}, h_i, \hat{p}(a), \alpha)$. Finally, to encourage exploration, we reset $R_t \leftarrow R_{t-1}$ even if the loss produced by $R_t$ is higher than the loss produced by $R_{t-1}$ with a decreasing probability over time, as shown in Line 21, Alg. \ref{alg:team} and as is typically done in simulated annealing procedures \citep{wang_gaining_nodate}.

\section{Empirical Evaluations}

We conduct empirical evaluations to explore whether and when \textsc{tr} and alternative AI advisors can be relied on to either add value or avoid diminishing the value offered by the human alone to their context. In particular, we assess AI advisors produced in practice and optimized to maximize their generalization performance, as well as AI advisors that are personalized based on factors produced in prior work. We further study what may be the risks, if any, from ignoring any of the pillars and why these risks arise. 
To gain further insights we analyze the distinct advising strategies that \textsc{tr}'s advisors produce relative to those of other advisors and examine how these strategies relate to an AI advisor's ability to reliably add value across contexts. These analyses offer novel insights into how an AI's personalized expertise and advising behaviors add value and avoid losses in a concrete context.

Similarly to prior work on  algorithmic frameworks for interactive contexts, establishing a method's behavior relative to alternatives requires establishing the robustness of the algorithm's performance in simulations across diverse contexts \citep{wanxueMS, madras_predict_2018, bansal_is_2021, clement2022, Dvijotham_Winkens_Barsbey_Ghaisas_Stanforth_Pawlowski_Strachan_Ahmed_Azizi_Bachrach_et, Frazer2024, pmlr-v162-verma22c}. Concretely, given that organizational trade-offs and human behaviors shape \textsc{tr} advisors' expertise and advising behavior, understanding AI advisors' reliability and relative performance requires assessment across settings in which humans exhibit diverse levels of expertise and discretion behaviors towards the AI advice and in contexts characterized by different decision-making domains and cost-benefit trade-offs. 
Below, we outline alternative AI advisors, the settings in which they were evaluated, and evaluation measures of an AI advisor's added-value.


\textbf{Advising Frameworks.} 
We consider AI advisors used in practice and which are optimized to maximize their generalization performance, as well as AI advisors that are personalized based on factors produced in prior work. We further study what may be the risks, if any, from ignoring any of the pillars and why these risks arise. Because prior work has not considered frameworks for producing AI advisors of which expertise and advising behavior are personalized to the human's discretion behavior, expertise, and the context's cost-benefit trade-off, we consider alternative AI advisors that are either not personalized or are personalized based only on subsets of the pillars. 
To allow attribution of differences in performance to the pillars of the framework we propose, all advisors use the same state-of-the-art method to generate the candidate rule sets of explainable rule-based AI advice and the same simulated annealing optimization framework to select a rule-set for advising the human \citep{wang2017bayesian}.

Below we outline each alternative and rationale for assessing its performance. In Appendix B we outline complete details for how each method selects and uses a rule set to advise the human. 

\begin{enumerate}
 \item \textbf{Task-Only Advisor (Current Practice)}  For every setting we consider, we produce advisors used in practice, which are trained to maximize their generalization performance and that advise the human on each decision. Comparison to such advisors allows us to explore the potential value-added of AI advisors widely employed in practice, across contexts.

\item \textbf{TR-No(ADB, Cost)}. We evaluate AI advisors that reflect an enhanced variant of prior works' consideration of the human's expertise \citep{when_advise_2023}. Specifically, \textsc{TR-No(ADB, Cost)} is personalized to improve decision outcomes by maximizing the AI's complementary expertise relative to that of the human and it advises selectively if the advice is expected to be superior to the human's independent decision. However, \textsc{TR-No(ADB, Cost)} advisors' expertise and advising behavior are not personalized to the human's algorithm discretion behavior nor the context's cost-benefit trade-off. Notably, while some prior work considered AI advisors that are personalized to individual's expertise, these methods were not designed to complement the human's imperfect discretion of AI advice nor did they consider the costs incurred by decision-makers when reconciling AI advice \citep{when_advise_2023}. 
We refer to this alternative as \textsc{TR-No(ADB, Cost)} to denote that it does not consider \textsc{tr}'s pillars of accounting for the human's ADB and the context's cost-benefit trade-off. 

\item \textbf{TR-No(ADB).} To assess the potential value across different contexts when an AI advisor is designed to offer advice that is assumed to be accepted by the human, we produce and evaluate AI advisors by accounting for all the pillars except for the human's ADB. To our knowledge, no prior work has developed or evaluated such AI advising methods that are personalized to the human's expertise and the context's cost-benefit trade-off in which decision-makers incur costs from reconciling AI advice. However, this framework builds on prior work that considered the human's expertise while assuming the human is an optimal advice-taker that incurs costs when advice is not provided \citep{bansal_is_2021}. Assessing \textsc{TR-No(ADB)}'s performance aims to understand the potential differential value across contexts from the learning and consideration of the human's ADB. 


\item \textbf{TR-No(Cost).} To our knowledge, no prior work had proposed to produce AI advisors whose expertise and advising behaviors are shaped by both the human's expertise and ADB. However, we evaluate \textsc{TR-No(Cost)} to assess the potential value of advisors that do not integrate the context's advising cost-benefit trade-off, but which include all other pillars -- i.e., AI advisors' expertise and selective advising  are personalized to the human's expertise and discretion behavior so as to maximize the \emph{decision} outcomes. 



\end{enumerate}

\begin{table}[!htbp]

\caption{Summary of advisor-generating methods, human expertise and discretion behaviors, and measures of AI-advisors' performances}

\begin{subtable}{0.9\textwidth}
    \centering

\makebox[\textwidth]{ 
\resizebox{0.6\textwidth}{!}{
\begin{tabular}{c|c|ccc}
\hline
\textbf{}                                                              & \textbf{}                                                                & \multicolumn{3}{c}{\textbf{ Accounts for:}}                                                                                                                                                            \\ \hline
\textbf{\begin{tabular}[c]{@{}c@{}}Advising \\ Framework\end{tabular}} & \textbf{\begin{tabular}[c]{@{}c@{}}Advises \\ selectively\end{tabular}} & \textbf{\begin{tabular}[c]{@{}c@{}}Human \\ expertise\end{tabular}} & \textbf{\begin{tabular}[c]{@{}c@{}}Human discretion \\ (ADB)\end{tabular}} & \textbf{\begin{tabular}[c]{@{}c@{}}Context's cost-benefit \\ trade-off\end{tabular}} \\ \hline
\textbf{\textsc{TR}}                                                   & \checkmark                                                               & \checkmark                                                           & \checkmark                                                     & \checkmark                                                                  \\
\textbf{\textsc{TR-no(ADB)}}                                           & \checkmark                                                               & \checkmark                                                           &                                                                & \checkmark                                                                  \\
\textbf{\textsc{TR-no(Cost)}}                                          & \checkmark                                                               & \checkmark                                                           & \checkmark                                                     &                                                                             \\
\textbf{\textsc{TR-no(ADB, Cost)}}                                     & \checkmark                                                               & \checkmark                                                           &                                                                &                                                                             \\
\textbf{Task-Only (Current Practice)}                                  & \textbf{}                                                                &                                                                      &                                                                &                                                                             \\ \hline
\end{tabular} }}
\end{subtable}
\subcaption{Advisor Generators}
\label{tab:adv_frameworks}

\hfill

\begin{subtable}{\textwidth}

\centering
\makebox[\textwidth]{ 
\resizebox{0.9\textwidth}{!}{%

\begin{tabular}{l|>{\raggedright\arraybackslash\sloppy}p{5cm}|>{\raggedright\arraybackslash\sloppy}p{5cm}|>{\raggedright\arraybackslash\sloppy}p{5cm}|>{\raggedright\arraybackslash\sloppy}p{5cm}|}
\cmidrule{2-5}
 & \multicolumn{2}{c|}{\textbf{Human expertise behaviors}} & \multicolumn{2}{c|}{\textbf{Human confidence and discretion behaviors (ADB)}} \\ \cmidrule{2-5} 
 & \textbf{Difficulty-Biased Decisions}  & \textbf{Group-Biased Decisions} & \textbf{Accuracy-Biased ADB} & \textbf{Group-Biased ADB} \\ \cmidrule{2-5} 
 & Likelihood of human error varies with task difficulty \citep{pmlr-v9-yan10a, wanxueMS}. & Likelihood of human error varies across types of tasks \citep{Adams_Buckingham_Lindenmeyer_McKinlay_Link_Marceau_Arber_2007,Wu_Gale_Hall_Dondo_Metcalfe_Oliver_Batin_Hemingway_Timmis_West_2016b}. & Human's discretion behavior is function of their confidence in their decision and AI's reported confidence \citep{will_you_accept}. Human's confidence in their independent judgment varies with but does not perfectly align with their true decision accuracy \citep{Klayman1999, JOHNSON2021203, CHONG2022107018}. & Human's discretion behavior is function of their confidence in their decision and AI's reported confidence \citep{will_you_accept}. Human confidence in their independent judgment varies across types of tasks \citep{Maserejian2009}. \\ \cmidrule{2-5}
    \end{tabular} }}
\end{subtable}
\subcaption{Simulated Human Expertise and ADB}
\label{tab:sim_behaviors}

\hfill

\begin{subtable}{\textwidth}
\centering

\makebox[\textwidth]{ 
\resizebox{\textwidth}{!}{
\begin{tabular}{@{}l|ll@{}}
\textbf{Metric}                                                                                                                                                                                                                                                                                                                                                                                                                                                                                                                                                                                                                    & \textbf{Definition}                                                                                                                         &  \\ \cmidrule{1-2}
Value Added                                                                                                                                                                                                                                                                                                                                                                                                                                                                                                                                                                                                                        & (Accuracy Improvement in TDL w.r.t. Human) $-$ (Cost-Benefit Trade-Off)*(Advising Costs Incurred)              &  \\
Accuracy Improvement w.r.t. Human                                                                                                                                                                                                                                                                                                                                                                                                                                                                                                                                                                                                 & $\frac{\text{(Number of Human Decisions Corrected Through Advising In Group)}}{\text{(Total Number of Decision Instances)}}$                                &  \\
Improvement in Team Decision Loss w.r.t. Human ($\uparrow$ TDL)                                                                                                                                                                                                                                                                                                                                                                                                                                                                                                                                                                                                & $\frac{\text{(Number of Human Decisions Corrected Through Advising In Group Weighted by Loss Incurred from Incorrect Decision)}}{\text{(Total Number of Decision Instances)}}$                                 &  \\
Advising Costs Incurred                                                                                                                                                                                                                                                                                                                                                                                                                                                                                                                                                                                                            & $\frac{\text{(Number of Pieces of Advice In Group)}}{\text{(Total Number of Decision Instances)}}$ &  \\
Advising Costs Incurred (In Accuracy Units [AU])                                                                                                                                                                                                                                                                                                                                                                                                                                                                                                                                                                                   & (Cost-Benefit Trade-Off)*(Advising Costs Incurred)                                                      &  \\
Advising Confidence \footnote{Note that the confidence of rule-based advice corresponds to the precision of the rule. Thus, while the confidence is a measure of the quality of the rules used for advising and estimated prior to deployment, advising accuracy is a measure of the quality of decision advice offered under selective advising. TR selectively offers a rule as advice only if the advice is expected to add value. Hence, rules with any given precision are invoked selectively based on their expected added value, and may yield higher or lower advising accuracy.}  & (Mean Confidence of Contradicting Advice for Instances Advised On In Group)                                                                &  \\
Advising Accuracy                                                                                                                                                                                                                                                                                                                                                                                                                                                                                                                                                                                                                  & $\frac{\text{(Number of Correct Pieces of Advice In Group)}}{\text{(Total Number of Pieces of Advice In Group)}}$               &  \\
Advising Rate                                                                                                                                                                                                                                                                                                                                                                                                                                                                                                                                                                                                                      & $\frac{\text{(Number of Pieces of Advice In Group)}}{\text{(Total Number of Decision Instances In Group)}}$                                   &  \\
Advice Acceptance Rate                                                                                                                                                                                                                                                                                                                                                                                                                                                                                                                                                                                                             & $\frac{\text{(Number of Pieces of Advice Accepted By DM In Group)}}{\text{(Number of Pieces of Advice Offered In Group)}}$                         & \\

Errors Avoided (\%)                                                                                                                                                                                                                                                                                                                                                                                                                                                                                                                                                                                                             & $\frac{\text{(Number of Decisions Corrected Through Advising in Group)}}{\text{(Number of Incorrect Independent Human Decisions in Group)}}$                         &
\end{tabular} 
}}

\end{subtable}
\caption{Measures of Performance}
\label{tab:measures}

\end{table}

\textbf{Data.} We use three data domains which in practice represent consequential decisions: the Heart Disease Dataset in which labels reflect the presence of heart disease for a patient \citep{misc_heart_disease_45}, the FICO dataset \citep{fico_2018} reflecting loan applications data used to inform loan decisions, and an HR employee attrition dataset in which labels reflect employee departure, and thus such data can be used in practice to inform employment retention decisions.\footnote{https://www.kaggle.com/datasets/pavansubhasht/ibm-hr-analytics-attrition-dataset, note this is a fictional dataset that represents a real-world task} 

\textbf{Simulations.} Similarly to prior works that developed methods for human-AI collaboration \citep{madras_predict_2018, bansal_is_2021, clement2022, Dvijotham_Winkens_Barsbey_Ghaisas_Stanforth_Pawlowski_Strachan_Ahmed_Azizi_Bachrach_et, Frazer2024, pmlr-v162-verma22c}, we explore AI advising performance for a wide range of settings by simulating diverse human expertise \citep{wanxueMS, madras_predict_2018, mozannar2020consistent, pmlr-v162-verma22c}, confidence, and corresponding ADB \citep{Klayman1999, JOHNSON2021203, CHONG2022107018, Maserejian2009} that builds on prior works on human behavior, summarized in Table \ref{tab:sim_behaviors}. Complete details of the simulation procedures and parameters are outlined in Appendix A.2. 

Following prior work, for each instance $i$ that a decision-maker undertakes, the human expertise is reflected by the  probability of the human correctly determining the label $y_i$; further, we evaluate AI advisors for human decision errors that are either (1) Group-Biased or (2) Difficulty-Biased \citep{madras_predict_2018, mozannar2020consistent, pmlr-v162-verma22c, wanxueMS}. \emph{Group-biased} errors reflect a decision-maker who exhibits a higher (lower) rate of error for a group of instances. For example, it has been shown that cardiologists have a higher likelihood of misdiagnosing a heart attack in women as compared to men \citep{Adams_Buckingham_Lindenmeyer_McKinlay_Link_Marceau_Arber_2007}. \emph{Difficulty-biased} reflects a well-documented phenomenon in which humans and AI  exhibit complementary strengths and weaknesses on decision-making tasks. Thus,  \emph{difficulty-biased} reflects how a human's accuracy varies across instances of different difficulty levels, where difficulty is measured from an AI perspective. Towards this, we use a task-only AI model's confidence scores as a proxy to determine whether an instance is difficult or easy for AI systems optimized for generalization performance on the task \citep{hendrycks17baseline}. Prior research has shown that AI systems optimized for generalization accuracy can achieve superhuman performance on cases that are easier for the AI than for a human to solve, for example because AI maintains consistent performance while human decision-makers are susceptible to contextual factors \citep{KarlinskyShichor2024}, such as fatigue and emotions \citep{Trinh2021-sc, Weinshall-Margel2011-yw, Danziger2011-mo}. Conversely, humans can perform better on instances that are challenging for AI, such as atypical (out-of-distribution) cases \citep{han_2021, CAO2024103910, KarlinskyShichor2024}. Such complementary performance patterns where humans and AI tend to make mistakes on different cases have been observed in high-stakes decision-making settings \citep{Dvijotham_Winkens_Barsbey_Ghaisas_Stanforth_Pawlowski_Strachan_Ahmed_Azizi_Bachrach_et}. The \emph{difficulty-biased} human behavior thus consists of a human who performs better on instances that AI finds difficult and worse on instances that AI finds easy.


The human's discretion of AI advice is reflected by the human's likelihood of accepting AI advice for a given instance. We build on prior work that modeled and validated human ADB in real human-AI  interactions \citep{will_you_accept}, and which found that humans are more likely to accept AI advice the higher the AI's confidence is relative to their own (possibly miscalibrated) self-reported confidence in their decision. We build on this work to map the human's self-reported confidence and the AI advisor's confidence to the human's likelihood of accepting the AI's advice. Similarly, following prior works' findings that humans do not have perfectly calibrated confidence in their decisions \citep{Klayman1999, JOHNSON2021203, CHONG2022107018},  we explore two types of human confidence behaviors. The first,  \emph{accuracy-biased ADB}, refers to individuals whose confidence in their own decision correlates with, but does not perfectly match, their  probability of being correct for a given decision instance.  \emph{Group-biased ADB} refers to when the human's confidence is determined by the group of individuals for which decisions are made. For example, a decision-maker may be very confident when making diagnoses for younger patients but less confident with respect to older patients. In Appendix A.2 we outline complete details of the different decision and ADB behaviors we explore.

\textbf{Measures of Performance.} 
\label{app:measures}
In a departure from prior work, we focus on the value that an AI advisor adds to its organizational contexts, accounting for the context-relevant trade-off between the cost of reconciling AI advice that conflicts with the human's judgment and its benefits to decision outcomes. Concretely, an AI advisor's added value is computed by the decrease in losses incurred by the human-AI team relative to the losses incurred by the human's independent decisions. 

We show the AI advisor's impact (decrease or increase) on the empirical Total Team Loss (TTL) from the expert and AI team's final decisions relative to the value produced by the expert's standalone decisions. TTL thus aggregates the loss from incorrect decisions, termed Team Decision Loss (TDL), and the cost incurred from reconciling contradictory AI advice, henceforth termed Advising Loss (AL): $\text{TTL} =\text{TDL} +\text{AL}$. We define the empirical Team Decision Loss (TDL) over a set of test data $D^*$ as follows: $\text{TDL} = \frac{{\sum}_{i=1}^{|D^*|}\lambda_0^{(1-y_i)}\lambda_1^{(y_i)}\mathbb{I}(y_i \neq \tilde{h}_i)}{ |D^*|}$, where $\lambda_0$ is the decision loss from false positives, $\lambda_1$ is the decision loss from false negatives,  $\tilde{h}$ is the DM's final decision after being advised, and $y_i$ is the ground truth correct decision. We consider both settings where $\lambda_0 = \lambda_1$ and $\lambda_0 \neq \lambda_1$. The total advising loss (AL) is given by: $\text{AL} = \frac{\alpha\sum_{i=1}^{|D^*|}|\hat{y}_i - h_i|}{|D^*|}$. Finally, to compute the value added by the advisor, we compare the TTL to the human's decision loss from their initial, independent decisions: $\text{HDL} = \frac{{\sum}_{i=1}^{|D^*|}\lambda_0^{(1-y_i)}\lambda_1^{(y_i)}\mathbb{I}(y_i \neq h_i)}{ |D^*|}$, such that $\text{Added Value} = \text{HDL} - \text{TTL}$. All evaluation metrics are summarized in Table \ref{tab:measures}. 
Note that the human decisions and ADB are stochastic. Hence, given a trained advising model, evaluation metrics are computed by averaging $50$ different repetitions of the human's initial and final decisions, $h$ and $\tilde{h}$, respectively.



In the next section, we first present results for contexts in which the cost of false positive errors is the same as that of false negatives, and later present results in an asymmetric loss context.


\section{Results}\label{sec:results}
\label{sec:results}

Below we first show results for the value added by different AI advisors across diverse contexts characterized by different task domains, human expertise and algorithm discretion behaviors, and contextual cost-benefit trade-offs. We then analyze and discuss differential expertise and advising strategies that  \textsc{ReV-AI} advisors give rise to to add value and avoid losses in two different contexts. 

\subsection{The value added and lost by AI advisors.}  
\begin{figure}[t!]
    \centering
    \makebox[0pt]{
    \includegraphics{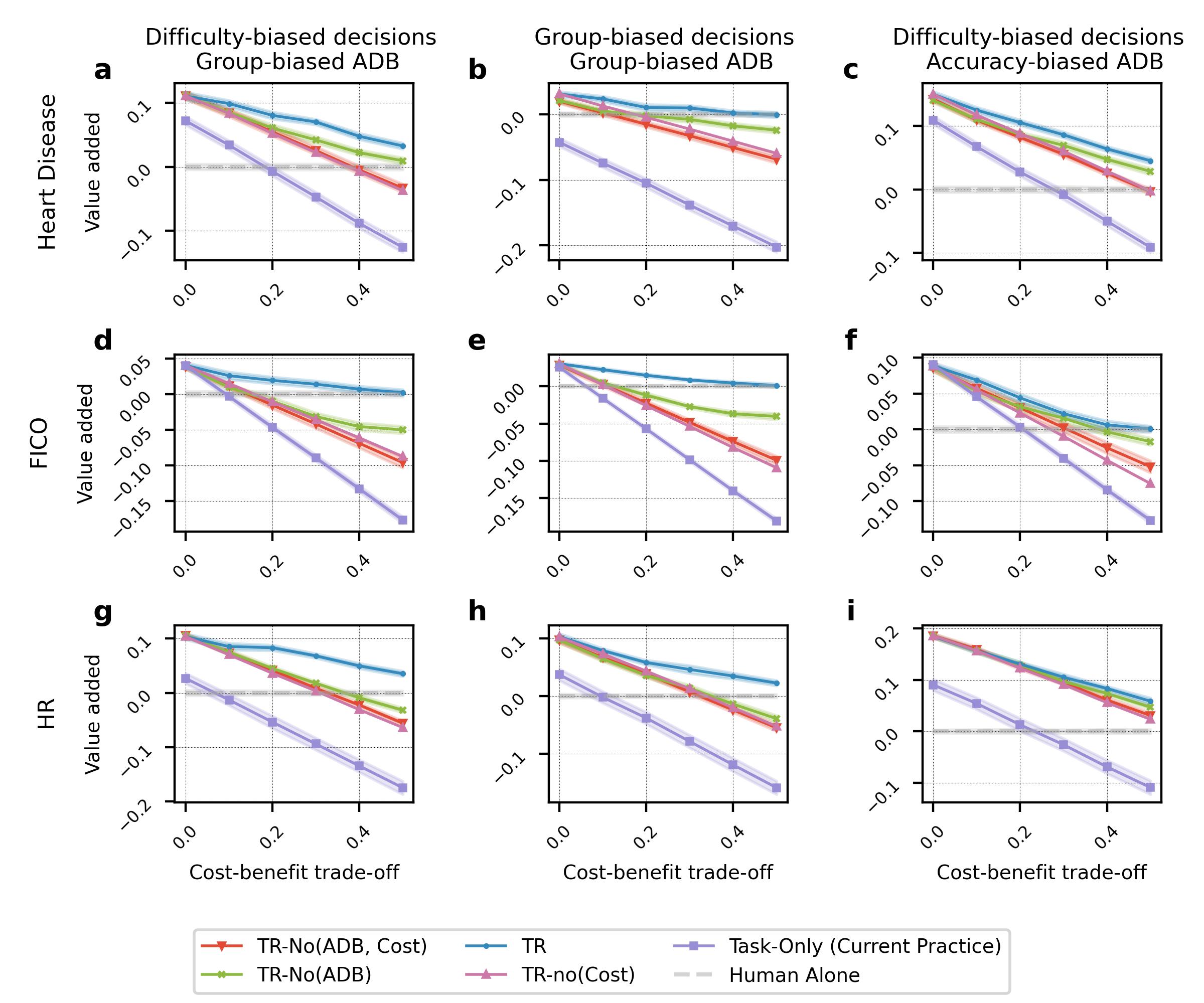}}
    \caption{Value added by different AI advisors relative to independent human for environments with varying trade-offs between advising costs and decision benefits for Heart Disease, FICO, and HR datasets. X-axis reflects contexts in which the cost from 1/x human engagements are equivalent to the loss from one incorrect decision: higher x-values reflect that fewer human engagements are permissible to achieve a given decision benefit. \textsc{task-only} model generalization accuracy is 0.78 (Heart Disease), 0.71 (FICO), 0.79 (HR). The human's standalone generalization accuracy for each dataset and decision behavior is 0.71 (Heart Disease Difficulty-Biased), 0.83 (Heart Disease Group-Biased), 0.73 (FICO Difficulty-Biased), 0.70 (FICO Group-Biased), 0.70 (HR Difficulty-Biased), 0.71 (HR Group-Biased). Thus, plots a,c,e,f,g, and h show settings in which a \textsc{task-only} model can achieve superhuman accuracy, while plots b and d show settings in which a \textsc{task-only} model cannot reach human performance. Results show average value-added +- SE (shaded region) over 10 repetitions.}
    \label{fig:main}
\end{figure}

For diverse contexts characterized by different data domains, human expertise and algorithm discretion behaviors, Figure \ref{fig:main} shows the value that different AI advisors add (or diminish) by advising the human (y-axis) as a function of the context's cost-benefit trade-off (x-axis). Note that for \textsc{TR} and \textsc{TR-No(ADB)}, each data point on each curve corresponds to a different AI advisor induced based on the human's behavior \emph{and the context's trade-off}. For each plot in Figure \ref{fig:main}, the figure label outlines the accuracy achieved by the human alone and by the model class (\textsc{task-only}) alone.  

In Figure \ref{fig:main}, the cost incurred from reconciling one disagreement with the AI is equivalent to the loss from $x$ decision errors. Thus, higher x-axis values correspond to contexts in which fewer human engagements to reconcile a disagreement are tolerated to achieve a given improvement in decisions.
In each plot, the dashed horizontal line is the value produced by the human alone (when the human makes decisions independently, without AI counsel). Thus, an AI advisor that produces a value above the horizontal line is adding value to its context beyond what the human is capable of alone, while a value below the horizontal line reflects that the AI advisor diminishes the value offered by the human alone.

All AI advisors struggle to add value when opportunities to improve the human's decisions are limited by the human's strong expertise or high confidence in their own judgment, both of which are characteristic of expert settings. When the human's expertise is strong, it is challenging to learn superhuman decisions that complement the human's already strong expertise; similarly, when the human is highly confident in their own judgment, even when such confidence is misplaced, it is more difficult to offer advice that is simultaneously superior and sufficiently convincing to the confident human. As an example, in Figure \ref{fig:main}b, the model class's generalization accuracy (0.78) does not match that of the human's (0.83) and the \textsc{task-only} advisor always diminishes value, even for contexts that incur \emph{no costs} from human engagement (x-axis value of $0$), demonstrating the model class's difficulty of offering sufficiently superior and convincing advice to improve the human's final decisions. Yet, \textsc{TR}, produced with the same model class, either adds value, and otherwise avoids the losses incurred by alternative advisors in contexts with low tolerance for the human's costly engagement. Similarly, in Figures \ref{fig:main}b, and d, we see that \textsc{TR} advisors produced with a model class that cannot match the human's accuracy reliably add value to their contexts and otherwise avoid diminishing the context's value. Furthermore, even when the AI model class achieves superhuman accuracy, shown in Figures \ref{fig:main}a,c,e,f,g, and h, all other advisors can diminish the value offered by the human alone, while \textsc{TR} advisors add the most value and do not diminish the human's value in all these contexts.

Overall, our results demonstrate that producing AI with superhuman performance is not enough to add value, nor is it necessary to add value. Specifically, AI advisors that do not advise selectively or are not informed by one or more of the pillars can fail to add the most value and risk diminishing the value achieved by the human alone, even when the AI surpasses the human's generalization accuracy. \textsc{Task-only} advisors, which are commonly deployed in practice, incur the greatest losses, even when the AI's accuracy exceeds that of the human expert. These results reveal the risks associated with the most prevalent AI advisor and are consistent with evidence from high-stakes fields such as radiology, in which AI advisors that exhibit performance comparable to or better than human experts were deemed cost-ineffective \citep{lebovitz_engage_2022}. In contrast, \textsc{TR} advisors can add value even when the model class cannot match human performance, and they effectively avoid diminishing the value offered by the human alone.

\subsection{TeamRules' distinct advising strategies.}

\begin{figure}[t!]
    \centering
    \makebox[0pt]{
    \includegraphics[]{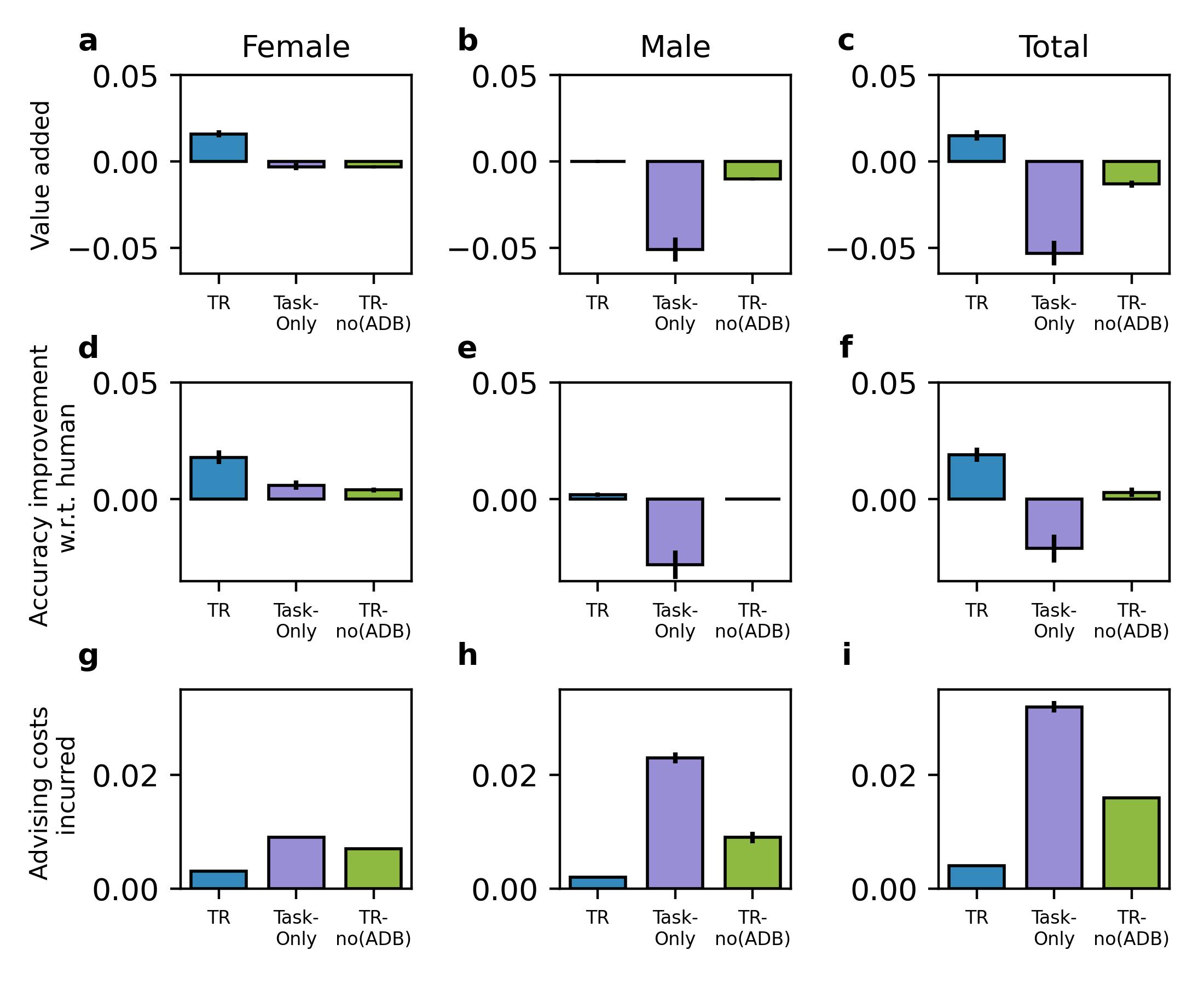}}
    \caption{Case Study 1 - Advising Outcomes,  defined in Table \ref{tab:measures}. \textsc{TR}, \textsc{task-only}, \textsc{TR-No(ADB)} advisors advise an over-confident expert. The context's trade-off is $0.1$, reflecting that reconciling up to 10 contradictory pieces of AI advice is deemed cost-effective if the contradictions yield at least one improved decision. Advising costs incurred shown are converted to units of decision loss given the trade-off (AU). The expert has an independent decision accuracy of 90\% on the Male population (majority), a 60\% accuracy on the Female population (minority), and an overall accuracy of 87.5\% on the entire population. This expert is highly confident (97.5\% confidence) on the entire population. Results are averages over 20 repetitions $\pm$ SE (vertical line at center of each bar).} 
    \label{fig:outcome_bars}
\end{figure}

\begin{figure}[t!]
    \centering
    \makebox[0pt]{
    \includegraphics[]{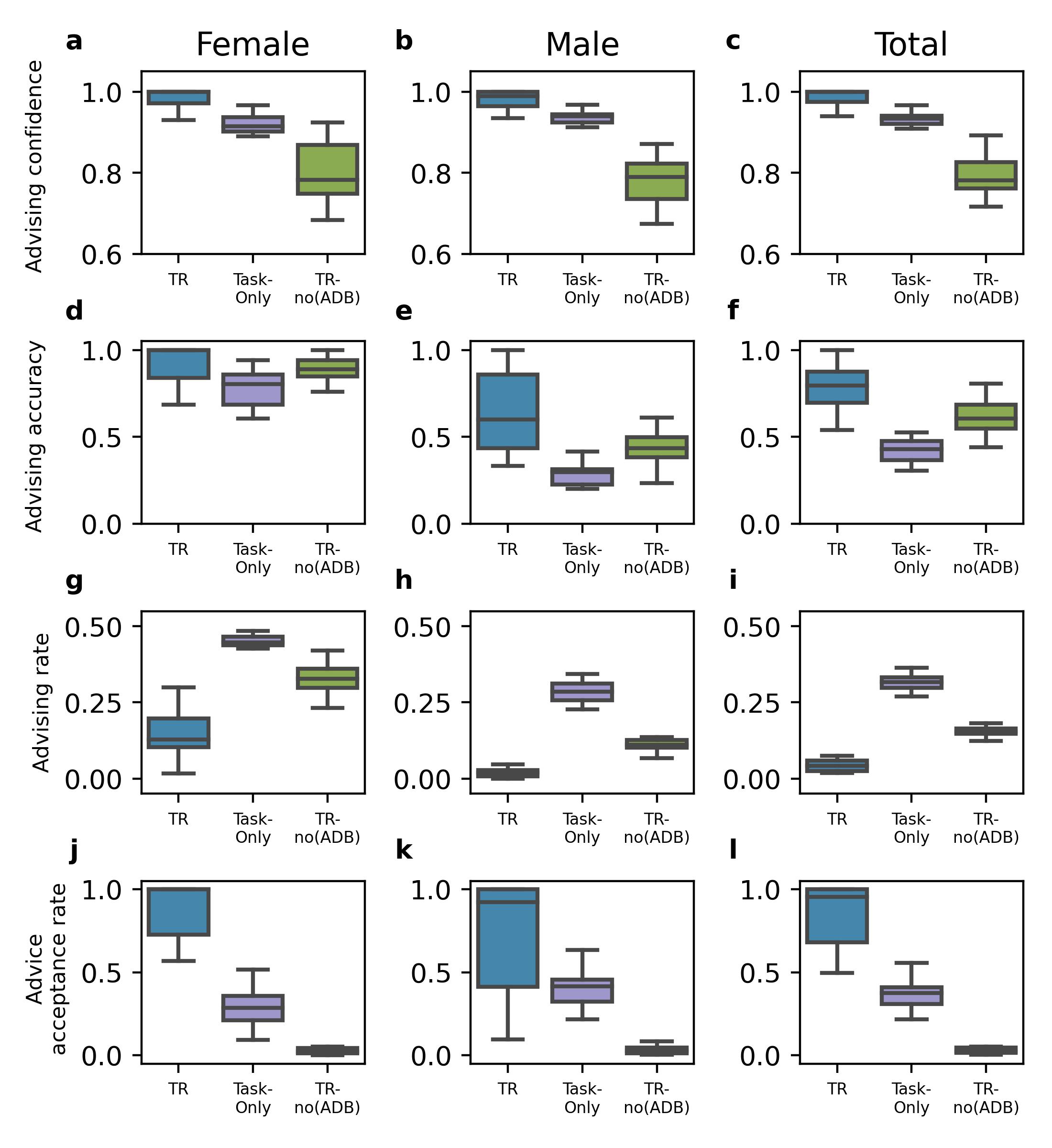}}
    \caption{Case Study - Advising Strategies: Properties of advice corresponding to Figure \ref{fig:outcome_bars}. Advising overconfident expert leads TR to learn and selectively offer higher accuracy and more confident advice that the expert is more likely to accept, leading to more improvements in decisions from fewer expert engagements. Results from 20 repetitions. AI advice corresponds to advice that disagrees with the expert's judgment and thereby can change the expert's decisions. All measures defined in Table \ref{tab:measures}.} 
    \label{fig:conf_boxes}
\end{figure}

We now analyze two case studies to explore \emph{how} \textsc{TR} advisors add value or avoid losses. Specifically, we examine how \textsc{TR}'s learned advice and advising behaviors differ from those of alternatives and how these impact their observed differential performance. Towards this, we use two Heart Disease diagnostic case studies, each characterized by different human decision making and algorithm discretion behaviors and different cost-benefit trade-offs. 

\subsubsection{Case Study 1: Advising a Biased and Overconfident Expert}

We utilize the Heart Disease diagnosis task \citep {misc_heart_disease_45}, and a context inspired by evidence of physicians  misdiagnosing heart failure in women at a higher rate as compared to men \citep{Maas2010, Wu_Gale_Hall_Dondo_Metcalfe_Oliver_Batin_Hemingway_Timmis_West_2016b, greenwood2018}. We simulate an expert with the reported performance who is also highly confident in their diagnoses for both men and women. While such an expert can benefit from an AI advisor which can improve their decisions on women in particular, it is challenging to produce advice that is sufficiently convincing to the expert. This is because such advice must not only be more accurate than the expert's, but also sufficiently more confident than the expert’s (miscalibrated) confidence in order to increase its likelihood of being accepted. The overconfidence of the expert limits the potential to improve their decisions in a cost-effective manner. 

Figure \ref{fig:outcome_bars} shows the outcomes when each AI advisor is paired with the expert to produce decisions and Figure \ref{fig:conf_boxes} shows the advising expertise and strategies produced by AI advisors. Importantly, we report the AI advisor’s \emph{advising} rate, accuracy, confidence, and acceptance rate. Unlike traditional evaluations of an AI’s autonomous decision performance, these measures capture the effective pathway through which AI advice impacts value—specifically when it contradicts human judgment and thus has the potential to alter outcomes. Metrics such as advising accuracy and acceptance rate pertain only to cases where the AI’s advice could change the human's decision, as opposed to instances of agreement between the human and AI. These measures therefore characterize the AI's expertise and advising behaviors through which AI advice either adds or diminishes value. To characterize \textsc{TR}'s distinct advising strategy we contrast it with that of the \textsc{task-only} and \textsc{TR-No(ADB)} advisors. 

As shown in Figures \ref{fig:outcome_bars}, \textsc{TR} adds the most value (Figure \ref{fig:outcome_bars}a,b,c) and it does so by improving more of the expert's final decisions  (Figure \ref{fig:outcome_bars}d, e,f) through fewer engagements with the human (Figure \ref{fig:outcome_bars} g, h, i). 
We see that \textsc{TR}'s distinct advising strategy is shaped by the pillars, including the expert's overconfident ADB, decision-making expertise, and context's trade-off. Given the expert's near perfect decision accuracy and high confidence for male patients, the scope for cost-effective advising on men is limited.  Hence, to complement the expert, \textsc{TR} advises highly selectively relative to the alternatives (Figure \ref{fig:conf_boxes}g,h,i). In particular, while most of the patients are males, 65\% of \textsc{TR}'s advice focuses on female patients, for whom the expert's decision accuracy is the lowest. \textsc{TR}'s advice for female patients ultimately adds the most value compared to alternatives (Figure \ref{fig:outcome_bars}a). While other advisors lead to large losses from cost-ineffective advising on male patients,  \textsc{TR}'s avoids these losses, and even leads to a small improvement in decisions for men.

The nature of \textsc{TR}'s advice is also distinct. \textsc{TR} learns and advises with more accurate (Figures \ref{fig:conf_boxes}d, e, f) and confident advice (Figures \ref{fig:conf_boxes}a, b, c) relative to alternatives, including particularly for women. Hence, not only does the expert's weak accuracy in diagnosing female patients offer opportunities for an AI advisor to add value, but \textsc{TR} also specializes to increase its complementarity on women by learning and offering more accurate and confident advice on female patients. Consequently, \textsc{TR}'s advice for female patients makes the overconfident expert nearly twice more likely to accept \textsc{TR}'s advice compared to \textsc{task-only}'s advice (Figures \ref{fig:conf_boxes}j, k, i), thereby improving more of the expert's decisions in a cost-effective manner. The \textsc{Task-only} advisor engages the human more often (Figure 5h) with contradictory, yet, less accurate (Figure 5f) and less confident (Figure 5c) advice on both populations, ultimately diminishing  value (Figure 4c). Similarly, the \textsc{TR-No(ADB)} advisor, which learns and selectively offers advice that aims to complement human expertise and the context's trade-off, provides highly accurate advice (Figures \ref{fig:conf_boxes}d) for female patients, yet the advice is not sufficiently convincing to the expert and is accepted less often  (Figures \ref{fig:conf_boxes}j, k, l), undermining the advisor's ability to change the expert's final decisions cost-effectively and add the most value (Figures \ref{fig:outcome_bars}a, b, c) .


\begin{table}[t!]
    \centering
    \caption{Case Study - Advising outcomes and behavior under asymmetric error costs: The cost of missing a heart diseases is higher than that of incorrectly diagnosing the disease for a patient. The expert's advise-taking reflected by their ADB similarly reflects the reality of having asymmetric error costs. The expert struggles to correctly diagnose younger patients with heart disease (60\% accuracy), while is effective at diagnosing patients that do not have the disease or are older (95\% accuracy). In the table, AI advice corresponds to advice that disagrees with the expert's judgment and thereby can change the expert's decisions. All measures defined in Table \ref{tab:measures}. 
    Two-tailed paired sample t-test is used to determine significance of difference between \textsc{TR} and \textsc{task-only} Value Added outcomes. $**: p<0.01, *:p<0.05$}
    \includegraphics[width=\textwidth]{Plots/asym_case.png}
    \label{tab:asym_case_table}
\end{table}

\subsubsection{Case Study 2: Advising in an Asymmetric Cost Setting.}
We now consider a setting inspired by high-stakes contexts in which failing to detect a consequential outcome, such as a disease, is more costly than incorrectly detecting a disease. Concretely, the loss from a false negative is three times the loss from a false positive $(\lambda_0=1, \lambda_1=3)$. The asymmetric loss from incorrect decisions also impacts humans' advice-taking behaviors, compelling an expert to be less likely to accept AI advice that contradicts the expert's assessment of the presence of the disease and more likely to accept a contradictory AI advice suggesting the presence of the disease. We explore how the properties of this context impact \textsc{TR} advisors' expertise and advising behaviors. Note that the loss minimization objective for all advisors reflects the decision loss $\mathcal{V}(y, d) = \lambda_0^{(1-y)}\lambda_1^{(y)}\mathbb{I}(y \neq d)$ where $\lambda_0$ and $\lambda_1$ are set to match the asymmetric costs of the setting. Thus, all methods account for the differential consequences of errors. Given the human's complex expertise and ADB, we simplify the analysis of the advisors' strategies by setting the context’s trade-off to $0$, so that any AI advice is cost-effective as long as it improves decision outcomes. Thus, an AI advisor's added value is equivalent to an improvement in the decision outcomes (team decision loss). 

Inspired  by evidence that hypertension in young adults more often goes undiagnosed than hypertension in older patients \citep{Johnson2014-fj}, we simulate a human expert with high accuracy when diagnosing older patients and patients who do not have the disease, while exhibiting lower accuracy when diagnosing younger patients with the disease. As noted above, the expert’s ADB is similarly affected by the higher false negatives costs.  Concretely, the expert's ADB is gender group-biased and modulated by the expert's propensity to accept AI advice based on the differential risks of different errors. More specifically, when the AI suggests a patient has the disease contrary to the expert's initial assessment, the likelihood of acceptance is increased by a factor of $1.5$.\footnote{The likelihood of acceptance is constrained such that it cannot exceed 1.} Conversely, if the expert initially assesses that the patient has the disease but the AI advises otherwise, the acceptance likelihood is reduced by a factor of $0.5$. 

Table \ref{tab:asym_case_table} shows the added value of AI advice, advising accuracies, advising rates, and advising confidence for \textsc{TR} and \textsc{task-only} advisors.  To complement our prior results we focus on highlighting \textsc{TR}'s differential expertise and advising strategy in this setting relative to those of the \textsc{task-only} advisor; however, we note that, consistent with other settings, \textsc{TR} adds more value than \textsc{TR-No(ADB)} in this setting.


Across all patients, we see that \textsc{TR} adds more value than the \textsc{task-only} advisor. Note that this context values improvement in decision outcomes at any cost for the expert, and preventing the expert from missing the disease in patients yields the most significant improvements in value. Hence, to add the most value an AI advisor should (1) excel at detecting the presence of the disease, (2) produce advice on the presence of heart disease that is convincing to expert, and (3) choose to offer this advice to the expert so as to avoid the costly mistakes. We see that in this context \textsc{TR} produces an AI advisor that exhibits this expertise and behavior. \textsc{TR} adds more value by specializing to yield near-perfect advising accuracy (0.982) for patients \emph{with} the disease-- i.e., on $98.6\%$ of the cases in which \textsc{TR} disagrees with the expert's judgment for patients in this group, \textsc{TR} recommends diagnosing the disease that was missed by the expert. As a result, \textsc{TR} adds 2.7 times the value offered by the \textsc{task-only} advisor for patients with the disease (0.218 vs. 0.081).

To impact the expert's final decision, the AI advice ought to not only contradict the expert's most costly errors, but also be sufficiently convincing to the individual expert so that its advice is accepted sufficiently often, given the expert's particular ADB. Indeed, given the expert is more likely to miss the disease in young patients, we see that \textsc{TR}'s advice is also sufficiently convincing to the expert for these patients, reversing $70.6\%$ of the expert's costly misses on young patients with heart disease. In contrast, the \textsc{task-only} advisor only helps the expert correct $49.2\%$ of their misdiagnoses in this group, $30\%$ fewer corrections than when working with \textsc{TR}. Consequently, \textsc{TR} adds 45\% more value than the \textsc{task-only} advisor for its advising on younger patients with the disease.  

Given the asymmetric loss, both the \textsc{task-only} and \textsc{TR} advisors over-diagnose patients without the disease. While \textsc{TR}'s losses from over-diagnosing are higher, given the asymmetric costs, \textsc{TR} selectively advises the expert almost exclusively to change negative diagnoses to positive, and rarely attempts to convince the expert to do the opposite. \textsc{TR}'s added-value is thus enabled by offering advice that effectively convinces the expert to avoid more of the costly expert errors for patients with the disease at the cost of a smaller loss from leading the expert to incorrectly diagnose some healthy patients. Further, both \textsc{TR} and \textsc{task-only} lead the expert to make more errors overall because both methods try to over-diagnose due to the assymmetric cost. However, while \textsc{TR} leads the expert to make $17.6\%$ more errors, all almost exclusively stemming from over-diagnoses of patients without the disease, \textsc{task-only} convinces the expert to make $41.4\%$ more errors overall, with some of those errors leading the expert to misdiagnose patients with the disease, leading to high costs and ultimately not producing as much value overall. 

While \textsc{TR} almost always advises that a patient has the disease when it provides advice, it is important to note that \textsc{TR} advises selectively. \textsc{TR} selects which cases are likely to have the disease, and withholds advice otherwise. While both \textsc{TR} and \textsc{task-only} provide recommendations that contradict the experts initial judgment on a similar percentage of cases overall ($0.279$ and $0.285$ for \textsc{TR} and \textsc{task-only}, respectively), \textsc{TR}'s recommendations are more valuable. Additionally, while \textsc{TR} generally provides advice with lower confidence than \textsc{task-only}, \textsc{TR}'s advice is sufficiently convincing to the expert, and, \textsc{TR} offers more confident advice on patients that have the disease than on patients that do not have the disease, further helping the expert determine when the recommendations should be followed appropriately. 

Finally, note that there can be superior or simpler AI strategies than those produced by \textsc{TR} in any given context. However, an effective and reliable method for learning AI advisors ought to \emph{autonomously} learn AI advisors' expertise and advising behaviors which, \emph{across} diverse experts and contexts, can be relied on to add value and avoid losses relative to alternatives. 

\subsection{Additional Results}

We present several additional analyses in the appendix that explore complementary aspects to our main findings. 

In Appendix C.1 we explore settings with very low tolerance for any added human engagement. Concretely, to add value in this context, the human’s decisions must be improved in at least \emph{one of every two cases} in which the AI engages the human with contradictory advice. These results aim to explore extremely challenging contexts, such as emergency medicine, in which experts’ time is an especially
scarce and costly resource. It is therefore highly challenging for an AI advisor to add value by offering cost-effective AI advice, given that more than 50\% of the times that advice is offered it ought to be simultaneously superior and accepted by the expert. This challenging setting increases the risk that AI advice offered would diminish the context's value. We find  that \textsc{TR} effectively  mitigate losses incurred by other AI advisors in such settings and is also able to add some value in some cases by specializing to this context. 



Appendix C.2 presents a second case study with an expert that exhibits appropriate confidence calibrated to their abilities (high confidence for male patients where the expert's accuracy is high, low confidence for female patients where accuracy is lower). While humans have been shown to have miscalibrated confidence in their own decisions, calibrated experts are better advice-takers (they are more likely to accept better advice and reject bad advice) and thereby this represents setting in which it is easier for AI advisors to add value. We aimed to explore the relative performances of different AI advisors in this context, as well as the settings in which different AI advisors are more likely to benefit or harm outcomes. The results show that, in this setting, \textsc{TR} produces an advisor that adds value by developing an advising strategy that complements the calibrated human expert. With the expert’s improved ADB, \textsc{TR} offers more frequent advice for female patients given the human's ADB decreases the risk that low quality advice is accepted. Furthermore, as the expert is not overconfident, superior AI advice need not be exceptionally confident to be accepted, thereby increasing opportunities, which \textsc{TR} leverages, to add value.

While \textsc{TR} assesses the value of offering each individual decision advice, in any high-stakes setting, it is advisable that AI advisors' overall expected added-value and safety are assessed \emph{prior} to deployment in a risk-free setting. This is important to reduce the risk of diminishing the value produced by the human alone. In Appendix C.3 we assess the effectiveness of this simple validation in contexts in which it is increasingly difficult to learn the human's ADB. Specifically, in the empirical results reported in this section, an imperfect model of the expert's ADB is learned from data. However, in Appendix C.3, we artificially add noise to the discretion model's inferences so that the model's estimations increasingly diverge from the human's true ADB, thereby undermining \textsc{TR}'s performance. We observe the expert's ADB on a risk-free validation set of tasks to empirically assess the added value of each AI advisor produced by \textsc{TR}, so that an advisor is not used to advise the expert on any of the decisions if it is not expected to add value over the validation set. We first find that, as expected, \textsc{TR}'s added value decreases with greater model-expert misalignment. We also find that \textsc{TR}'s inherent mitigation of losses from AI advice remains highly effective,  and that a validation-based robustness mechanism improves safety via slightly better mitigation of value diminishment. 

\section{Related Work}\label{sec:Related}


Prior work had not proposed or empirically studied a value-based framework for producing AI advisors in AIaDM settings with the goal of adding the most value and mitigating losses and which accounts for the pillars through which AI advice may add or diminish value in AIaDM contexts. Concretely, most prior research that developed AI methods for human-AI collaborative decision-making had considered deferral collaborations in which each decision is either handled \emph{autonomously} by the AI or deferred to the human,  thereby obviating the need to consider an AI \emph{advisor} to a human and how AI advice impacts outcomes \citep{madras_predict_2018, gao2023learning}. 

Earlier work on AI advisors did not focus on high-stakes settings and did not propose or share insights on methods that integrated the proposed pillars through which AI advice can add or diminish value in high-stakes AIaDM settings \citep{bansal_does_2021}. Our empirical evaluations include enhanced versions of these advisors. Specifically, \cite{bansal_is_2021} assume the human is a perfect advice-taker, such that the human always rejects advice that is expected to diminish value and always accepts value-adding advice, thereby obviating the need to learn the human's imperfect ADB and assess its impact on decision outcomes. Yet, humans are idiosyncratic and imperfect advise-takers, and their imperfect discretion of AI advice is a key pathway through which AI advice may fail to add the most value or introduce losses. The AI advisor developed in  \cite{bansal_is_2021} also offers advice on each decision instance. In the empirical evaluations, \textsc{TR-No(ADB)} is an enhanced version of this advisor which aims to complement the human's expertise but also selectively offers advice only if it is estimated to add value given a fixed ADB that assumes the expert will accept all AI advice, thereby accommodating an ADB that is not learned from historical data.  

\cite{clement2022} considered a sequential decision making setting in which each decision impacts the likelihood of transitioning into a subsequent state. Hence, this work does not consider AIaDM contexts in which a human handles a set of independent decisions, and is not directly applicable to the contexts we focus on in this work.  Conceptually, \cite{clement2022}  do not focus on high-stakes contexts and do not advocate for value-based advisors that integrate the \textsc{ReV-AI} pillars. \cite{clement2022} considers the human's likelihood of accepting AI advice, however, contrary to evidence from prior work, assume that the human's likelihood to accept advice in a given state is fixed and independent of properties of the advice itself, such as the advice's confidence. Consequently, this work also does not aim to  produce advice that is convincing to the human advisee so as to be more likely to be accepted.

\cite{when_advise_2023} proposed an AI advisor trained on only on the task  and independently of the human advisee and paired with a policy that predicts whether the AI advisor is more likely to be accurate than the human on any given task; thus, unlike \textsc{ReV-AI}, this work did not aim to produce advisors whose expertise and persuasiveness complement the human's decision-making weaknesses and ADB so as to add the most value. However, the advice of their task-only advisor is offered when it is expected to be more accurate than the human’s independent decision.  In the empirical evaluations, \textsc{tr-No(ADB, Cost)} is an enhanced version of this advisor which selectively offers advice if it is estimated to be more accurate than the human's judgment.


 

As we demonstrated in this work, reliably adding the most value across contexts hinges on methods that holistically consider and offer a response to the pathways impacting the value-added of AI advice in high-stakes decision-making contexts, such as how the advice itself impacts the human's likelihood of accepting it, and potential losses from excess human engagement to reconcile consequential differences with the AI advice. Our analyses reveal the differential expertise and advising behaviors that can arise in \textsc{ReV-AI} advisors relative to alternative AI advisors, and how these properties are instrumental to adding the most value. We also demonstrate the opportunity costs and loss of value when AI advisors are not designed to reliably add the most value. To our knowledge, no prior work has offered these insights.

Our work builds on research that documented  outcomes of human-AI collaboration in high stakes decision-making contexts  \citep{cai_human-centered_2019,BALAGOPAL2021102101, Rezazade_Mehrizi2023-ia, Jacobs2021-zs, NBERw31422, Heterogeneity_Salz_Agarwal_Rajpurkar_2024, soares_fair-by-design_2019, green_disparate}.  A recent meta-analysis by \cite{Vaccaro2024} finds that AI assistance does not consistently enhance outcomes across domains and tasks. For example, \cite{Heterogeneity_Salz_Agarwal_Rajpurkar_2024} found that AI-assisted radiology diagnostics helped some experts while harming others. Notably, this body of work considers AI models corresponding to the \textsc{task-only} benchmark in our work.  Yet, these works do not discuss how such inconsistent outcomes may arise. Our empirical studies offer complementary and novel insights into how AI advice in AIaDM contexts may lead to losses or yield differential value across different humans and contexts. Our work also highlights AIaDM contexts with the highest risks of incurring the most significant losses from AI advice. Our work also shows that, across different human expertise, ADB, and context's tradeoffs, the \textsc{task-only} advisor is more likely to introduce losses, and that it introduces the greatest losses among all the alterative AI advisors.

The \textsc{ReV-AI} framework can be used to develop \textsc{ReV-AI} advisors based on different model classes that are preferred in a given context. We use the \textsc{\normalsize{h}\scriptsize{y}\normalsize{rs}} model class \citep{wang_gaining_nodate} to produce and evaluate AI advisors and to develop a concrete, proof-of-concept \textsc{ReV-AI} method.  \textsc{\normalsize{h}\scriptsize{y}\normalsize{rs}} produces inherently interpretable rules and was developed to offer partial interpretability for black-box models.  Rule-based models learn sparse rule sets composed of if-then statements, thereby providing inherently interpretable and faithful rationale for its recommendations \citep{rules_why/15-AOAS848, vc_rules, rudin2019stop,wang2021scalable,wang2018multi}.

\section{Limitations and Future Work}\label{sec:conclusion}

Recent work has revealed significant and consequential differences in the benefits of AI advice for different decision makers and contexts with different trade-offs between the costs and benefits resulting from AI advice. Our work advocates for a fundamental shift towards a value-based paradigm for the development and evaluation of AI advisors that \emph{reliably} maximize their added value by design. This shift requires new frameworks to produce AI advisors that are shaped by contextual factors that impact their added-value and reliability in avoiding losses.  We propose a framework for producing \textsc{ReV-AI} advisors for AIaDM settings, defined by the form of human-AI collaboration shown in Figure \ref{fig:dec_sequence_intro}, that future work can build on to develop new \textsc{ReV-AI} frameworks for other forms of human-AI collaborations. For a given form of preferred human-AI collaboration, there can be opportunities to produce \textsc{ReV-AI} advisors that enhance the value they add to their context and mitigate losses. Following the  \textsc{ReV-AI} framework for a given human-AI collaborative setting entails identifying key pathways through which the AI collaborator may either add or diminish value and how it moderates the means to achieve outcomes. 

Our work develops a proof-of-concept method for producing rule-based advisors yielding inherently interpretable advice, which has been shown to be beneficial and preferred in practice for advising high-stakes decision-making \citep{rudin2019stop,wang2021scalable,rich,BALAGOPAL2021102101,edit_machine, pmlr-v139-biggs21a, Subramanian_Sun_Drissi_Ettl_2022, prescriptive_relu}. Yet, it would be valuable for future studies to explore and advance \textsc{ReV-AI} frameworks to propose \textsc{ReV-AI} advisors for additional model classes. Relatedly, our method considers classification decision tasks, which are representative of many consequential decisions in practice, including diagnostic decisions and loan approvals, which we consider in the empirical evaluations. We hope that our work will spark new research on \textsc{ReV-AI} advisors for other kinds of consequential expert decisions, such as treatment choices.

Advancing algorithmic approaches for high-stakes interactive contexts entails establishing the robustness of their performance across diverse contexts and diverse expert decision-making and advice-taking behaviors in simulations  \citep{wanxueMS, madras_predict_2018, bansal_is_2021, clement2022, Dvijotham_Winkens_Barsbey_Ghaisas_Stanforth_Pawlowski_Strachan_Ahmed_Azizi_Bachrach_et, Frazer2024, pmlr-v162-verma22c, editorial-bus-ds}. 
While our simulations are grounded in established human behavior patterns from prior research, they may not fully capture the complexity of human experts in practice. Given our results show significant promise, it would be valuable for future work to study \textsc{ReV-AI} advisors' value-added when paired with diverse human experts across diverse contexts. Importantly, while evaluations with human experts, such as physicians, are highly costly and less accessible, involving actual experts conducting high-stakes decisions would be insightful given experts have been shown to exhibit distinct properties such as strong task expertise and corresponding high confidence. These distinct properties can give rise to AI advising behaviors and outcomes that differ from those produced by advisors for laypersons.

Similarly to most innovations, the introduction of \textsc{ReV-AI} advisors for high-stakes experts can have a variety of long-term implications leading to further progress but also possibly new challenges that future research can identify and study. Our value-based approach provides a foundation for investigating possible effects through extended field studies.


\section{Conclusions and Managerial Implications }\label{sec:disc}
Although AI advising tools used to advise experts can outperform individual experts in some tasks, they may not add value and instead incur losses. These outcomes may harm high-stakes practices, erode trust in AI, and undermine progress. Yet, no prior work has offered a framework  outlining  pathways through which such harms can arise, nor offered applicable means to produce AI advisors that generate the most value across AIaDM contexts. This work sheds a timely light on these challenges. 

To enhance the positive impact of AI advisors, we advocate for a shift towards a value-driven perspective. Central to this approach is the objective to shape AI advisors to offer the most added-value, and the characterization of key pillars -- pathways through which AI advice can either add to or detract from value in practice and that underlie the development of \textsc{rev-ai} advisors that can be relied on to add the most value and mitigate losses. The framework we introduce can be applied directly to different contexts and can be adapted to different model classes to produce \textsc{Rev-AI} advisors. Across different decisions that an expert handles, the \textsc{Rev-AI} advisor's expertise, the persuasiveness of their offered advice, and the choice to selectively offer any learned advice are shaped by the advisee and context to yield the most value and avoid losses. 

The empirical results we share also suggest why the focus on producing superhuman AI  is not enough to add value, nor necessary.  Concretely, our empirical results reveal that \textsc{Rev-AI} advisors can add more value even when the underlying model class cannot match the expert's decision-making performance. We also find that the most prevalent task-focused AI advisor in practice trained to yield the best decisions autonomously may not add value or introduce harms and diminish value across contexts, even if the underlying model class yields superhuman performance. 


We explored diverse contexts, including particularly ones in which we find AI advisors struggle to add value and risk diminishing value - such as when the human's confidence in their own judgment is highly miscalibrated, when the human's expertise is strong relative to a general AI, and when the context is less tolerant of expert's engagement with AI advice to improve decision outcomes. Our results show that across such contexts, \textsc{Rev-AI} advisors consistently add more value than alternatives, while the most prevalent AI advisors in practice may not add value and introduce the greatest risk of diminishing value from their contexts. Our case studies also reveal that to achieve these outcomes \textsc{Rev-AI} advisors give rise to distinct and advantageous properties that are complementary to their context: \textsc{Rev-AI} advisors exhibit not only (1) more advantageous expertise that better complements the expert's abilities but (2) simultaneously produce advice with ADB-bound information that is sufficiently persuasive to complement the human's ADB, and (3) they offer advice that is more likely to warrant the human's engagement so as to add the most added-value. 




Our results on the differential value offered by AI advisors deployed in practice and by \textsc{Rev-AI} advisors suggest that there may be significant unrealized value that diverse AIaDM contexts could gain from producing and deploying \textsc{Rev-AI} advisors. More value can be gained using existing model classes that are preferred in any given context, even in contexts where the preferred model class cannot match the human expert's performance.

This work also suggests that to facilitate AI's positive impact on practice, it is crucial for organizations to appropriately assess the value that AI-advising systems produce, accounting for costs that have largely been unaccounted for in existing methods for developing AI advisors, such as any additional time costs incurred by decision-makers from reconciling AI advice. This is significant both to uncover failures and reasons for abandonment of existing AI advising tools (e.g.,  \cite{lebovitz_engage_2022}) but, crucially, to facilitate the development and adoption of value-creating AI advisors. 

In current practice, AI tools are often developed and evaluated based on their performance for a cohort of experts. Yet, AI advisors which may harm outcomes of individual decision-makers may not be acceptable in certain high-stakes settings \citep{Heterogeneity_Salz_Agarwal_Rajpurkar_2024}. Furthermore, even in contexts where such outcomes are acceptable, AI advisors that can reliably add the most value to all individual experts they advise would further increase the value added for any cohort.

Designing \textsc{Rev‑AI} advisors entails up front costs for tailoring them to the decision‑maker’s expertise, ADB, and the specific trade‑offs of the operating context. Given decision outcomes in high-stakes are consequential, the upfront training cost in which human experts interact with the system in a risk‑free (offline) environment are unlikely to exceed the long-term gains after deployment; still, this can vary across applications. As in current practice, we recommend an incremental roll out: deploy \textsc{Rev‑AI} at limited scope to establish that the benefits from initial training outweigh the costs before scaling up.

As we stand at a critical juncture in the integration of AI into high-stakes human decision-making settings, this work offers not a final solution, but a principled framework on which future research and practical application can build on to benefit high-stakes decision-making. By centering on value rather than performance alone, we hope to shift the conversation from what AI can do to what it should do—be mindful of the human expert, the context, and the stakes involved. While much remains to be explored, we believe that \textsc{Rev-AI} advisors illuminate a practical and responsible path forward—one that invites both further innovation and careful reflection in equal measure.

\bibliographystyle{unsrt}

\newpage
\bibliography{sample}  
\newpage
\appendix
\section{Experiment Details} \label{exp_deets}
Here we provide full details on datasets and simulation procedures used for all experiments. 
\subsection{Heart Disease, FICO, HR Datasets}
Instance counts for all datasets are shown in Table \ref{tab:data_counts}.

\begin{table*}[h]
\centering
\caption{Dataset Instance Counts}
\label{tab:data_counts}
\begin{tabular}{l|lll}
\textbf{Dataset} & \textbf{Train Instances} & \textbf{Validation Instances} & \textbf{Test Instances} \\ \hline
Heart Disease & 505   & 87  & 127   \\
FICO          & 6,120 & 801 & 1,080 \\
HR            & 568   & 38  & 143  
\end{tabular}
\end{table*}

\subsection{Human Behavior Simulation Details}
These details correspond to the results shown in Figure \ref{fig:main}.

\subsubsection{Step 1: Simulating Human Decisions}
\label{app:sim_dec}
Following prior work which simulated human decision-behavior for human-AI collaborative settings \citep{madras_predict_2018, mozannar2020consistent, pmlr-v162-verma22c}, we represent human decision behavior as the probability of the human correctly determining the label $y_i$ for any instance $i$. Specifically, we create two types of decision behaviors: (1) Difficulty-Biased and (2) Group-Biased. 

\emph{Difficulty-Biased} behavior is simulated by first training an out-of-the-box probabilistic model, such as logistic regression, to predict the correct label for unseen task instances. We can denote the probabilistic output of the model as $\hat{p}_{logistic}(y|x)$, and we use this output as a proxy for the difficulty of the instance for an AI model. AI tends to under-perform relative to humans on out-of-distribution instances, and out-performs humans on in-distribution instances \citep{han_2021, CAO2024103910}, and the logistic regression output can be used to assess the extent to which an instance is in- or out-of-distribution \citep{hendrycks17baseline}. We thus refer to instances that are more likely to be out-of-distribution as more ``difficult" for an AI. Specifically, we denote the proxy value for instance $i$'s difficulty with $d_i$, such that $d_i = 2|\hat{p}_{logistic}(y=1|x) - 0.5|$. If $d_i > d_j$, then we consider instance $i$ as \emph{easier} to classify than instance $j$, because the probabilistic model predicts its class with a higher likelihood (is more confident of the instance's label). Finally, we partition the instance space such that each value $d_i$ for all instances $i$ maps to the simulated human's likelihood of being correct, and we create regions such that the human has a high accuracy for some difficult instances, and a low accuracy for some easier instances. Specifically, we set the human's accuracy to $60\%$ whenever $d > d_t$ and $100\%$ whenever $d <= d_t$, where $d_t$ is a dataset specific difficulty threshold. $d_t$ is set to $0.6$ for the Heart Disease dataset, $0.3$ for the FICO dataset, and $0.8$ for the HR dataset. Values were chosen so that the human would have $100\%$ accuracy on a smaller part of the dataset (roughly 25\%-30\% of instances) and 60\% on the rest, leading each human to have an overall accuracy ranging from 70\%-80\%.  

\emph{Group-Biased} behavior is simulated by selecting a feature upon which to partition the instance-space. For example, in practice, some doctors have been shown to exhibit gender and age bias when making heart disease diagnostic decisions \citep{Adams_Buckingham_Lindenmeyer_McKinlay_Link_Marceau_Arber_2007}. Simulating group-biased behavior thus involves selecting a feature that partitions the space into groups, such as gender, and subsequently setting the simulated DM's likelihoods of being correct to different values for each partition. Such a partition allows us to simulate expert decision-making biases often seen in practice. For each dataset, if an instance meets a feature-based \emph{condition}, the human's accuracy is set to 60\%, otherwise, it is set to 95\%. The conditions for each dataset are as follows: 

\begin{itemize}
    \item Heart Disease: Age $< 50$
    \item FICO: Number of  Satisfactory Trades $< 24$
    \item HR: Age $> 32$
\end{itemize}

\subsubsection{Step 2: Simulating Human Confidence in Their Decisions} 
Next, given a simulated human decision-behavior, we simulate human confidence behaviors. A decision-maker (DM) with \emph{accuracy-biased} confidence exhibits low confidence in decisions they are less likely to get correct and high confidence on decisions they are more likely to get correct. Specifically, the accuracy-biased human has confidence $c^H_i = p(h_i = y_i) \pm  \kappa$, meaning their confidence equals the likelihood of making a correct decision, offset by a constant value. This offset term ensures that the human's probability of being correct on a given instance is not exactly equal to their confidence, as humans rarely have perfectly calibrated confidence in their decisions \citep{Klayman1999, JOHNSON2021203, CHONG2022107018}. We also generate \emph{group-biased} DM confidence behavior. Humans with group-biased confidence behavior have confidence that varies with both instance difficulty and along a selected feature. For instance difficulty, we use the same difficulty values and thresholds $d_t$ described in Section \ref{app:sim_dec}. The features selected for each dataset are \emph{gender} for the Heart Disease and HR datasets, and \emph{external risk estimate score} for the FICO dataset. For all datasets, when $d < d_t$, confidence is set to $0.9$. When $d >= d_t$, we use the following group-based rules: 

\begin{itemize}
    \item Heart Disease \& HR: Confidence is set to $0.2$ when \emph{gender} is male, otherwise confidence set to $1$. 
    \item FICO: Confidence is set to $0.2$ when \emph{external risk estimate} $< 65$, otherwise confidence set to $1$. 
\end{itemize}

\subsubsection{Step 3: Simulating and Modeling Human ADB}
To simulate humans' ground truth ADB, $p(a|c^M, c^H, \hat{y} \neq h)$, e.g., the likelihood with which they would accept contradicting AI advice, we follow prior work that explored various model forms of humans' ADB as functions of their self-reported (and possibly miscalibrated) confidences in their decisions and the AI advisor's reported confidence values \citep{will_you_accept}. Specifically, we utilize the Adjusted Na\"ive Bayes (ANB) CPT-Based utility component: 

\begin{align}
    & \hat{u}^t_{accept} = (1+\beta_{ADB})w(\frac{1}{1+\frac{(1-a) \dot (a-b)}{a \dot b}}) - \beta_{ADB} \\ 
    & \hat{u}^t_{reject} = 1-(1+\beta_{ADB})w(\frac{1}{1+\frac{(1-a) \dot (a-b)}{a \dot b}})
\end{align}

Where: 

\begin{align}
    & a = \frac{(c^M)^\gamma}{(c^M)^\gamma + (1-c^M)^\gamma} \\
    & b = \frac{(c^H)^\gamma}{(c^H)^\gamma + (1-c^H)^\gamma} \\
    & w(p) = \frac{(p^k)}{p^k + (1-p)^k}
\end{align}

The ANB CPT-Based utility component characterizes how the human DM estimates the utility of accepting and rejecting AI advice. We then utilize the basic Logit selection component:

\begin{align}
    p(a|c^M, c^H, \hat{y} \neq h) = \frac{\exp(\delta * \hat{u}^t_{accept})}{\exp(\delta * \hat{u}^t_{accept}) + \exp(\delta * \hat{u}^t_{reject})}
\end{align}

The selection component characterizes how the human DM stochastically decides whether to accept AI advice given their estimate of the utility provided from accepting AI advice. We selected the basic Logit selection component because applying human-based adjustment on the utility component
of the two-component model always helps to increase the model’s
predictive performance if the basic selection component was Logit \citep{will_you_accept}. Additionally, utilizing the basic Logit component rather than the human-adjusted component allows us to make fewer model parameter assumptions. We select the ANB CPT-Based utility component because it was found to generally lead to models that best fit to human behavior across treatments and regardless of selection component choice \citep{will_you_accept}.

We utilize the following parameter values for our simulation of human ADB: $\delta = 5$, $k = 0.63$, $\gamma = 0.95$, $\beta_{ADB} = 0.5$. The $\delta$ parameter model reflects the human decision maker’s sensitivity
to their estimates of utility gained/lost from correct/incorrect decisions: when $\delta \rightarrow 0$, the human decision maker takes actions randomly; when $\delta \rightarrow \infty$, the human decision maker always takes the action with the optimal estimated utility. We select a value such that the human would generally take the action with optimal estimated utility but not always. The human determines their estimated utility from correct/incorrect decisions by transforming their confidence and the AI's reported confidence through several functions that rely on parameters $k, \gamma$, and $\beta_{ADB}$. 

The parameter $\gamma$ varies from 0 to 1 and controls the degree to which the human adjusts their own confidence values and the AI's reported confidence values to $0.5$ before integrating them into a single value that represents the human's confidence in the AI's decision. When $\gamma = 0$, all the probabilities are transformed to 0.5, while when $\gamma = 1$, no adjustment is applied to probabilities before the human integrates the confidences. We select $\gamma = 0.95$ to represent a human that slightly adjusts probabilities closer to 0.5. Once the human has an estimate of the AI's probability of being correct, this probability is further transformed into a perceived probability, motivated by human tendency to perceive probabilities non-linearly \citep{will_you_accept}. The $k$ parameter determines how the likelihood they believe the AI to be correct is transformed into a perceived probability. We select a value of $k = 0.63$, which corresponds to an inverse-S shaped probability weighting function, i.e. the human overweights extreme events, and underweights common events. We select this value because expert DM's tend to overestimate the likelihood of rare diseases (rare event) and thus underestimate the lack of rare diseases (common event) \citep{jamainternmed2021}. Finally, given the human's final perceived probability of the AI being correct, $w(p)$, the $\beta_{ADB}$ parameter determines how much utility they expect to gain from accepting/rejecting AI advice. The $\beta_{ADB}$ parameter can be used to increase the disparity between utility gained from decisions expected to be correct and incorrect. A $\beta_{ADB}$ value of 0 means the expected utility gained from accepting advice is equal to $w(p)$ and the expected utility gained from rejecting advice is equal to $1-w(p)$. As $\beta_{ADB}$ increases, the gap between expected utility gains increases, corresponding to higher stakes settings. We thus set $\beta_{ADB}$ to 0.5, slightly increasing the perceived stakes across all settings in our main results.

Recall that, given the human's ADB is unknown, the \textsc{ReV-AI} learning framework requires the collection of data and subsequently induces a model to estimate the human's likelihood of accepting AI's advice for any given instance: $p(a|c_M, c_H, \hat{y} \neq h)$. Following our framework, for a subset of decision instances we generate data reflecting the simulated human's independent decisions, confidence, and ADB in response to an AI recommendation and confidence. The AI recommendation and confidence is provided to the human initially by a logistic regression model trained on a withheld subset of data. In the experiments reported here, we then induce $\hat{p}(a|c^M, c^H, \hat{y} \neq h)$ from the interaction data between the human and the logistic regression model using an out-of-the-box implementation of XGBoost \citep{xgboost}.

\subsection{Training Parameters}
\label{app:train_params}
\begin{itemize}
    \item $T = 2000$
    \item $C_0 = 0.01$
    \item $\beta_0 = 0.05$
    \item $\beta_1 = 4$
    \item $\beta_2 = 10,000$
\end{itemize}

We follow prior work and established practices to set the maximum rule length $\beta_1$ and minimum support $\beta_2$ \citep{wang2017bayesian,wang2018multi,wei2019generalized}. Specifically, rule-based models utilize rules of lengths 2-4 and a minimum support threshold $\beta_2$ of 5\%. These broadly accepted standards are informed by consistent findings for settings that reduce the risk of overfitting and and promote intepretability. 

\section{Alternative Rule-Production Methods}
\label{app:alt_rules}

\subsection{Task-Only}

The \textsc{task-only} method, inspired by the Bayesian Rule Sets method \citep{wang2017bayesian}, produces a rule set that is optimized for maximum generalization accuracy on the task data and which covers all instances (provides advice on all instances). Accordingly, the \textsc{task-only} approach offers decision advice $\hat{y}_i$ given by: 

\begin{equation} \label{dec_rule_taskonly}
\hat{y}_i = 
   \begin{cases}
    1, & \text{if } \mathbb{C}(R, x_i) \\
    0, & \text{otherwise}
\end{cases} 
\end{equation}

And it offers model confidence $c^M_i$:

\begin{equation}
\label{conf_rule_taskonly}
\begin{aligned} 
c^M_i &=  \begin{cases}
\underset{r \in R : \mathbb{C}(r,x_i) = 1}{\max}\Bigg[\frac{\sum_{j=1}^{n}\big(\mathbb{I}\{\mathbb{C}(r, x_j) \wedge \mathbb{I}(y_j = \hat{y}_i)\}\big)}{\sum_{k=1}^{n}\mathbb{C}(r, x_k)}\Bigg], & \scriptstyle{\text{if } \hat{y}_i = 1}\\
\frac{\sum_{j=1}^{n}\mathbb{I}\{y_j = \hat{y}_i \wedge y_j = 0\}}{\sum_{k=1}^{n}\mathbb{I}(\hat{y}_i = 0)}, & \scriptstyle{\text{otherwise}}
\end{cases}
\end{aligned}
\end{equation}

The loss is given by: 
$$\mathcal{L}_{task-only}(y, \hat{y}) = \mathcal{V}(y, \hat{y})$$ where $\mathcal{V}(y, d)$ is the same context relevant loss utilized in the corresponding \textsc{TR} model. The full optimization algorithm used for the task-only approach is Algorithm 1 from \citep{wang2017bayesian}.

\subsection{TR-no(ADB, Cost), TR-no(ADB), TR-no(Cost)}
\label{app:TR-no}

The \textsc{TR-no(ADB, Cost)}, \textsc{TR-no(ADB)}, and \textsc{TR-no(Cost)} frameworks produce rule sets and provide advice for some instances, but may defer to the human if their advice is not expected to add value given the pillars they consider. The \textsc{TR-no(ADB, Cost)} approach accounts for human expertise, the \textsc{TR-no(ADB)} approach accounts for both human expertise and the context's advising cost-benefit trade-off, and the \textsc{TR-no(Cost)} approach accounts for both human expertise and ADB. All approaches are variants of \textsc{TR} and work similarly. Models that do not account for ADB use a fixed discretion model $\hat{p}(a|c^M, c^H, \hat{y} \neq h)$ as input. Whereas $\hat{p}(a|c^M, c^H, \hat{y} \neq h)$ is an estimate of the human's ground truth discretion behavior in \textsc{TR}, $\hat{p}(a|c^M, c^H, \hat{y} \neq h) = 1$ in \textsc{TR-no(ADB)} and \textsc{TR-no(ADB, Cost)}, meaning that these models are trained with the assumption that the human would accept all advice offered (does not consider their ADB). Next, the methods that do not account for the cost-benefit trade-off, \textsc{TR-no(ADB, Cost)} and \textsc{TR-no(Cost)} are optimized with $\alpha = 0$, mean they assume there is no cost to providing AI advice, regardless of the context.

\section{Additional Results}
\label{app:additional}
In this section, we provide results on additional settings and offer additional analysis on the benefits offered by \textsc{TR}'s mechanisms. Specifically, we demonstrate how \textsc{TR} can add more value than alternatives when advising costs are especially inhibitive ($\alpha > 0.5$) and when the decision loss incurred from false positives and false negatives is unequal. We provide an additional case study to assess how \textsc{TR} adds value relative to alternatives when an expert decision-maker has confidence that is calibrated to their decision-making strengths and weaknesses. 
Finally, we also demonstrate how the value-added by \textsc{TR} decreases as discretion model accuracy decreases, and we demonstrate how a simple robustness mechanism can ensure that advisors optimized using low quality discretion models will not diminish the value offered by the human alone. 

\subsection{Prohibitive Advising Costs}
\label{sec:high_costs}
\begin{figure*}[t!]
    \centering
    \makebox[0pt]{
    \includegraphics[width=0.7\textwidth]{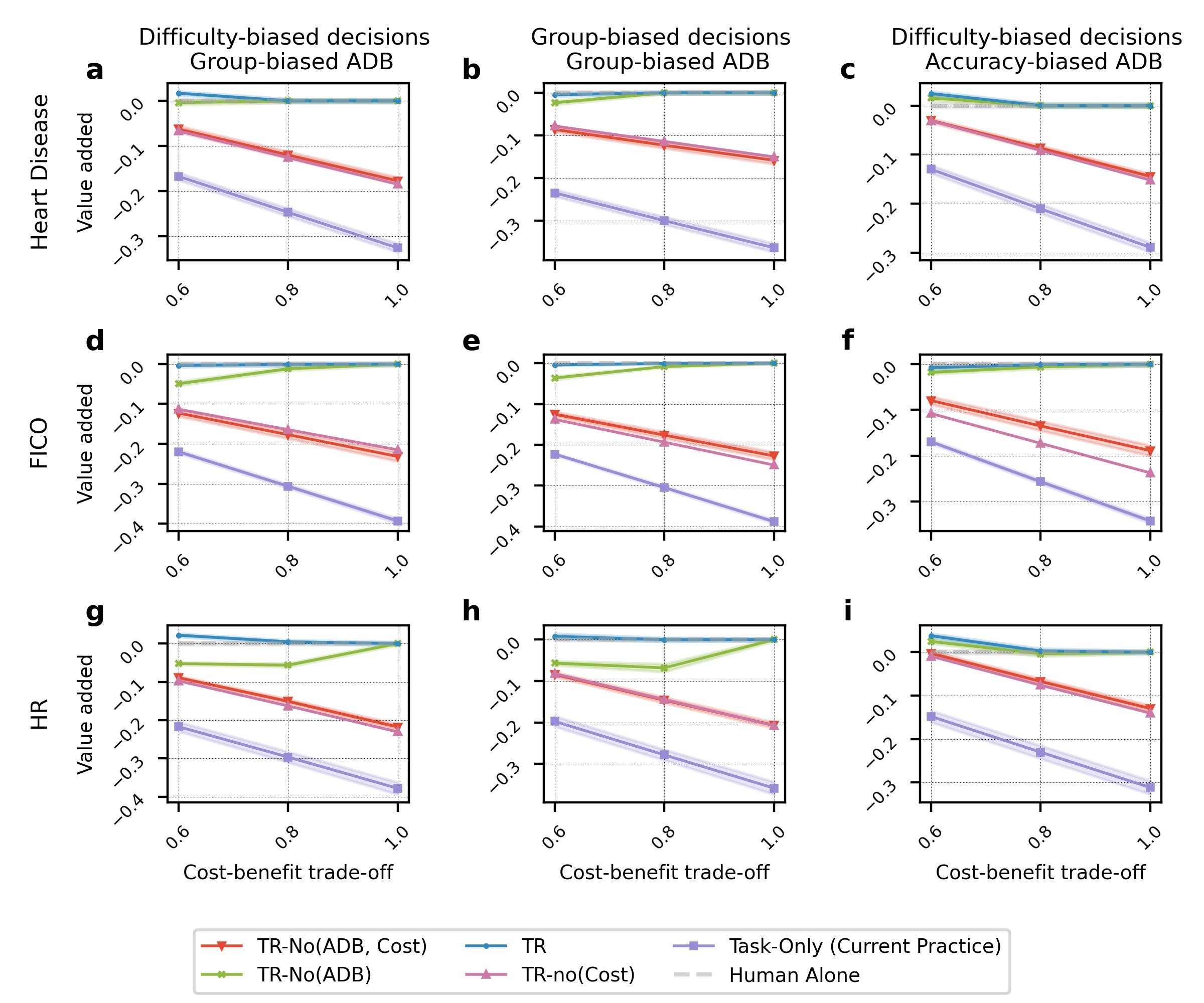}}
    \caption{Value added by AI advisors relative to human for environments with varying yet excessively inhibitive trade-offs between advising costs and decision benefits for Heart Disease, FICO, and HR datasets. X axis reflects contexts in which the cost from 1/x human engagements are equivalent to the loss from one incorrect decision: higher x-values reflect that fewer human engagements are permissible to achieve a given decision benefit. Results show average value-added +- SE (shaded region) over 10 repetitions.}
    \label{fig:high_cost}
\end{figure*}

In our main results, we show the value-added by \textsc{TR} relative to alternatives for $0 < \alpha \leq 0.5$. In this section, Figure \ref{fig:high_cost}, we show how \textsc{TR} can also add some value and does not diminish the value offered by the human even when the advising cost-benefit trade-off is especially inhibitive ($0.6 \leq \alpha \leq 1$). The alternative frameworks produce advisors that greatly diminish the human's standalone value across settings, while \textsc{TR} generally learns that it cannot add any value and refrains from advising the human. In some cases, it is even able to add some marginal value, such as in some settings when $\alpha = 0.6$.  

\subsection{Additional Case Study}
\label{sec:docB}

\begin{figure*}[t!]
    \centering
    \makebox[0pt]{
    \includegraphics[width=0.9\textwidth]{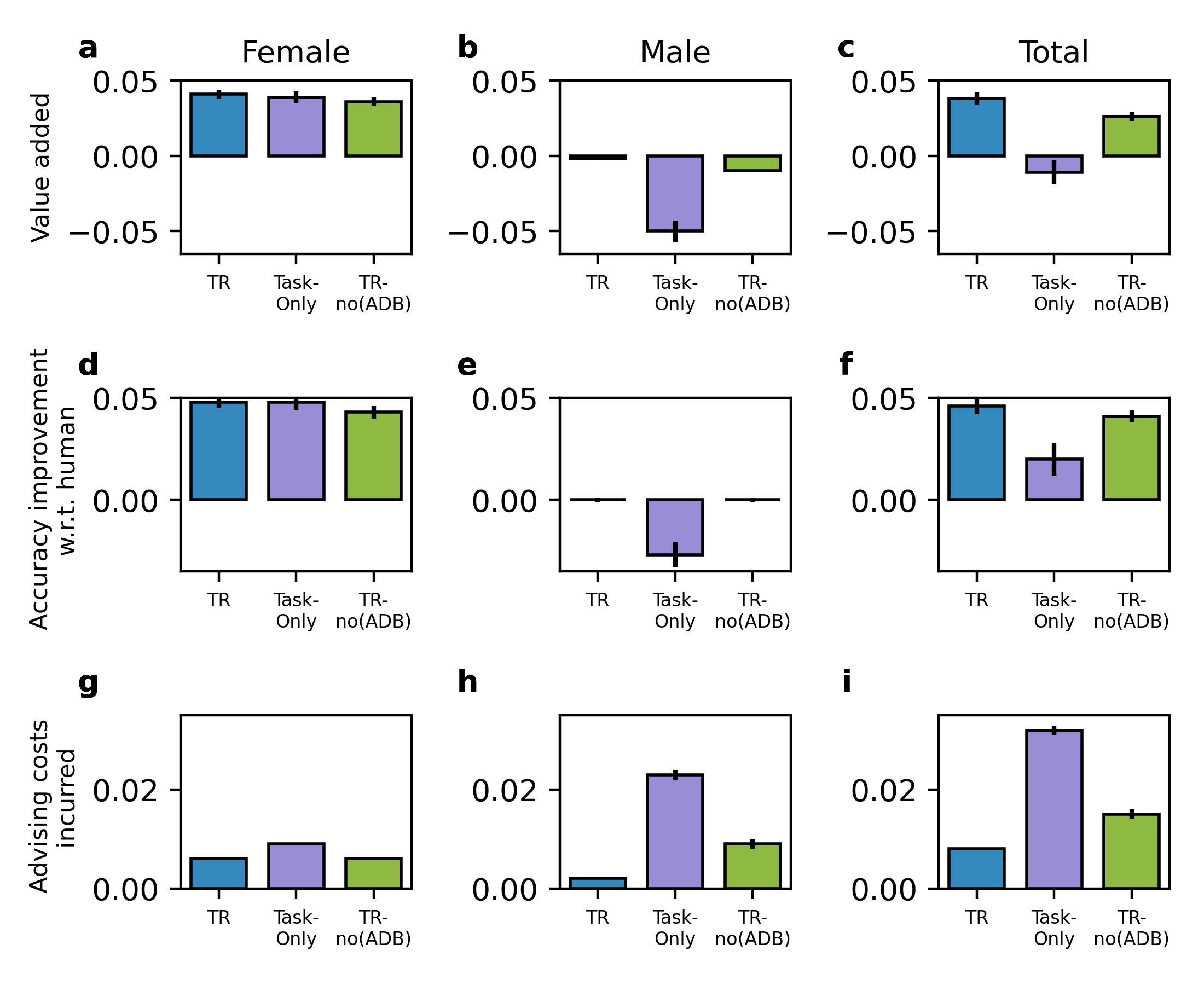}}
    \caption{Case Study - Outcomes: Advising outcomes when \textsc{TR}, \textsc{task-only}, \textsc{TR-no(ADB)} advisors advise an expert with relatively appropriate confidence given their expertise. The context's trade-off is $0.1$, reflecting that reconciling up to 10 contradictory pieces of AI advice is deemed cost-effective if the contradictions yield at least one improved decision. The expert has an independent decision accuracy of 90\% on the Male population (majority), a 60\% accuracy on the Female population (minority), and an overall accuracy of 87.5\% on the entire population. This expert is highly confident (97.5\% confidence) on the the male population and less confident (60.5\% confidence) on the female population. Results are averages over 20 repetitions $\pm$ SE (vertical line at center of each bar).} 
    \label{fig:outcome_bars_docB}
\end{figure*}

\begin{figure*}[t!]
    \centering
    \makebox[0pt]{
    \includegraphics[width=0.9\textwidth]{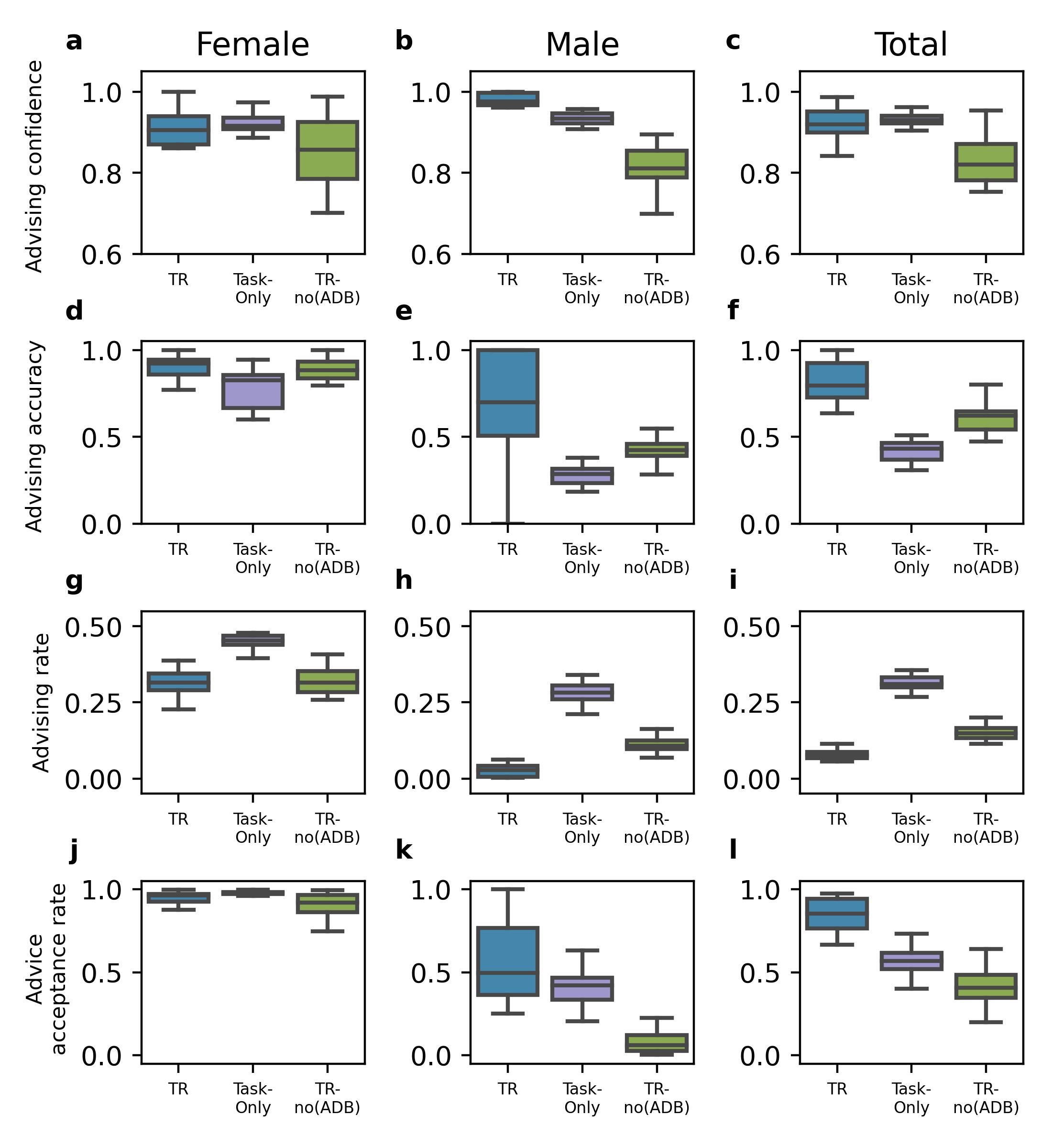}}
    \caption{Case Study - Advising Strategies: Properties of advice corresponding to Figure \ref{fig:outcome_bars_docB}. Advising an expert with relatively appropriate confidence in their abilities leads TR to learn and selectively offer high accuracy advice that is sufficiently convincing given the expert's confidence. Results from 20 repetitions. } 
    \label{fig:boxes_docB}
\end{figure*}

We simulate an expert, who we will refer to as Doctor B and who exhibits the same decision behavior as the expert in the main results section, who we will refer to as Doctor A. Doctor B has high accuracy when diagnosing male patients, and low accuracy when diagnosing female patients. However, in contrast to Doctor A, Doctor B is aware of their weakness when diagnosing female patients, and thus has appropriately low confidence (60\%) when diagnosing female patients while remaining slightly overconfident on male patients. Results are shown in Figures \ref{fig:outcome_bars_docB} and Figures \ref{fig:boxes_docB}.

We find that \textsc{TR} again focuses on providing majority of its advice for the female population, however, it adapts its advising strategy by producing lower confidence advice on the female population than when advising Doctor A. This allows for \textsc{TR} to offer more advice to Doctor B than to Doctor A, and because the threshold for providing cost-effective advice is lowered given Doctor B's low confidence and consequential receptiveness to advice, more decisions can be cost-effectively improved. \textsc{TR} also produces a different advising strategy when advising Doctor B on female and male patients. Doctor B is less receptive to advice for male patients, so \textsc{TR} offers less advice albeit with higher confidence on the male population relative to the female population. This strategy leads \textsc{TR} to add more value compared to the other advisors to Doctor B, significantly improving decisions cost-effectively on the female population without significantly diminishing the value Doctor B produces when diagnosing male patients. 

This case study represents a greater opportunity for adding value compared to the case study shown in the main results because Doctor B is receptive to advice on cases they would most benefit accepting advice for. Consequently, all advisors are able to add more value to this expert than they are able to add to Doctor A, however, \textsc{TR} still adds the most value to both experts relative to alternatives, demonstrating that \textsc{TR} is able to adapt its advising strategy to provide value-adding advice regardless of whether an expert's ADB is calibrated to their strengths and weaknesses. When opportunities to improve on the expert's behavior are limited, such as for the Doctor A, other methods cannot add value at all, while \textsc{TR} is able to add some value by offering only a few pieces of highly-confident and cost-effective advice. When there exist greater opportunities to improve on the expert's behavior, \textsc{TR} adds the most value by offering more advice cost-effectively than it is able to when opportunities are limited.

\subsection{Impact of discretion model accuracy on advising performance}

\label{sec:disc_bad}

\begin{figure}[t!]
    \centering
    \makebox[0pt]{
    \includegraphics[]{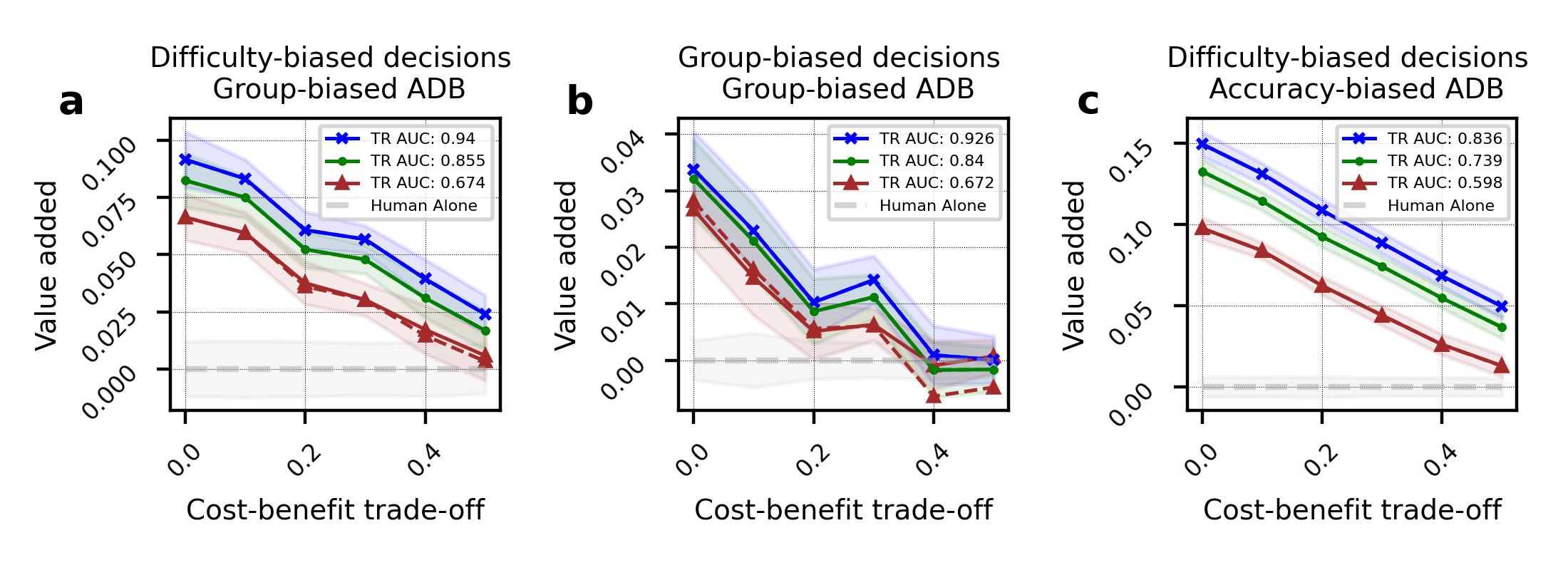}}
    \caption{Results when an already trained \textsc{TR} advisor is paired with different humans that exhibit ground truth ADB that deviates from the ADB \textsc{TR} was trained on. AUC values in the legend represent AUC of \textsc{TR}'s estimated model of the human's ADB on the human's ground truth ADB, such that lower AUC values mean the human's ADB deviates more from the model used to train \textsc{TR}. Solid lines show results of \textsc{TR} paired with the human while utilizing a robustness mechanism, while the dashed lines show the same pairing but without implementing the robustness mechanism. In many cases the lines overlap, meaning the mechanism does not have any effect. However, the robustness mechanism ensures that the advisor does not diminish the human's value, as shown in the middle panel.}
    \label{fig:disc_bad}
\end{figure}

In this section, we introduce and assess the benefit of a robustness mechanism that aims to further safeguard an AI advisor from offering advice that might diminish the human's own value. Concretely, the simple mechanism involves an empirical evaluation of the AI advisor's value implications when paired with the human on a validation set;  if the advisor is found empirically to diminish the human's value, the advisor is not used. In practice, assessing how the advisor performs with the individual expert on a set of risk-free  validation tasks is instrumental for producing reliable, value-adding advisors. 

\textsc{ReV-AI} advisors are designed to complement the human ADB by utilizing a discretion model which \emph{estimates} the human's ADB. In all main results and case studies, we simulate real-world procedures by first learning a discretion model from human-AI interaction data, which produces imperfect estimates of the human's true ADB. While imperfect, the discretion models produced in the main results were decent approximations of the human's true ADB (AUC $> 0.9$). In this section, we further analyze the value-added from \textsc{TR}'s  advice as the human's ADB deviates further from the discretion model used to optimize \textsc{TR}. It is critical to understand AI-advising performance under imperfect discretion estimates because, in practice, it may not be possible to learn an accurate discretion model. Further, it is critical to have mechanisms in place to ensure poor discretion models do not lead to advisors that can diminish human standalone value. 

We conduct an experiment on the Heart Disease dataset in which we pair a trained \textsc{TR} advisor with different simulated humans. The first human exhibits the same ADB that was used to train \textsc{TR}, and thus the AUC of the discretion model evaluated on the human's true ADB is $> 0.9$. For the subsequent humans we add random noise to their ADB such that the discretion model used to train \textsc{TR} is less and less representative of the human's true ADB. We include in the legend the AUC of the discretion model on each human's ADB. Lower AUC values mean the human exhibits a true ADB that deviates more from what \textsc{TR} was trained on. 

Results are shown in Figure \ref{fig:disc_bad}. The solid lines show model performance after a robustness mechanism using validation data was applied, while the corresponding dashed lines of the same color demonstrate the result when the robustness mechanism is not applied.\footnote{Note that in most cases, the dashed and solid line overlap, meaning the robustness mechanism did not activate to produce different results}  

Intuitively, we find that as the human's ADB deviates more from the model used to train \textsc{TR}, the value-added by \textsc{TR} diminishes. In some cases, we find that \textsc{TR} can begin to diminish the human's value if opportunities for improvement are limited (such as at higher costs) and the robustness mechanism is not applied, such as in Figure \ref{fig:disc_bad}b. With the robustness mechanism applied, the model would not be deployed if it diminishes the human's standalone value, preventing possible losses, and demonstrating how critical it is to evaluate AI advising models on low-stakes validation data and ensure mechanisms are in place to prevent possible losses.

\end{document}